\documentclass[a4paper,floats,floatfix,showpacs,11pt]{article}
\pdfoutput=1 

\usepackage{jheppub} 

\usepackage[normalem]{ulem}
\usepackage{xcolor}

\usepackage[T1]{fontenc} 
\usepackage{tikz-cd}
\usepackage{caption}
\usepackage{float}
\usepackage{subcaption}
\graphicspath{{figs/}}
\title{\boldmath A Unified Causal Framework for Nonlinear Electrodynamics Black Hole  from Courant-Hilbert Approach:  Thermodynamics and Singularity}

\author[a]{H. Babaei-Aghbolagh$^{\ast}$}
\author[b]{Komeil Babaei Velni$^{\ast}$}
\author[a,c]{Song He}
\author[b]{Fateme Isapour}

\affiliation[a]{Institute of Fundamental Physics and Quantum Technology, \& School of Physical Science and Technology,  Ningbo University, Ningbo, Zhejiang 315211, China}
\affiliation[b]{Department of Physics, University of Guilan, P.O. Box 41335-1914, Rasht, Iran}
\affiliation[c]{Max Planck Institute for Gravitational Physics (Albert Einstein Institute),
	Am M\"uhlenberg 1, 14476 Golm, Germany}
\emailAdd{hosseinbabaei@nbu.edu.cn}
\emailAdd{babaeivelni@guilan.ac.ir}
\emailAdd{hesong@nbu.edu.cn}
\emailAdd{fatemehisapour97@gmail.com}

\abstract{We develop a unified framework for analyzing black hole thermodynamics and spacetime structure in Einstein gravity coupled to causal nonlinear electrodynamics (NED) in asymptotically anti--de Sitter backgrounds. The electromagnetic sector is governed by a Generalized Nonlinear Electrodynamics (GNED) Lagrangian obtained from a root-$T\bar T$ deformation constructed via the Courant--Hilbert approach, ensuring both duality invariance and causal propagation. This theory contains ModMax, Generalized Born-Infeld (GBI), and self-dual logarithmic electrodynamics as continuous limits. Within this framework we obtain exact charged AdS black hole solutions and perform a detailed study of their thermodynamic properties, including mass, temperature, entropy, and free energy. The resulting phase structure exhibits van der~Waals-type transitions between small and large black holes and features a characteristic swallowtail in the free energy at the critical point.

We further investigate the internal geometry, showing that the nature of the central singularity is determined by the matter fields that source the spacetime. An analysis of the Kretschmann scalar reveals how mass and electric charge jointly govern curvature divergence in ModMax black holes.
We derive an explicit charge-to-mass bound within a causal logarithmic electrodynamic theory. This extends the finite self-energy property of the point charge beyond the standard Born-Infeld model. This bound cleanly distinguishes black holes from naked singularities, showing that naked singularities occur precisely when the mass parameter is smaller than the electromagnetic self-energy of the point charge. This provides a clear energetic criterion for horizon formation in the absence of a Cauchy horizon.
}

\begin{document}
\maketitle

\begingroup
\renewcommand{\thefootnote}{\fnsymbol{footnote}}
\setcounter{footnote}{0}
\footnotetext{Corresponding authors: Komeil Babaei Velni; H.~Babaei-Aghbolagh .}
\endgroup

\flushbottom

\section{Introduction}\label{0} 
Black holes are a fundamental prediction of Einstein's general relativity, whose theoretical existence was anticipated long before observational confirmation~\cite{Penrose:1964wq,Chandrasekhar:1985kt}. Within the Einstein--Maxwell framework, which unifies general relativity with classical electrodynamics, black holes can possess electric charge alongside mass and angular momentum~\cite{Misner:1973prb}. The Kerr--Newman solution is a prominent exact solution describing a rotating, charged black hole in this theory~\cite{Newman:1965tw}. More comprehensive solutions exist, such as the Plebański--Demiański family, which incorporates additional parameters, including NUT charge and uniform acceleration, thereby providing one of the most general classes of exact solutions in Einstein--Maxwell theory~\cite{Plebanski:1976gy}.

		Black holes in asymptotically anti-de Sitter (AdS) and de Sitter (dS) spacetimes exhibit fundamental differences from their asymptotically flat counterparts. In AdS backgrounds, topological black holes emerge with event horizons that may possess positive, zero, or negative constant curvature~\cite{Birmingham:1998nr}. The AdS geometry, characterized by a conformal boundary at spatial infinity and a globally timelike Killing vector, permits a consistent definition of conserved quantities such as mass, electric charge, and angular momentum~\cite{Henneaux:1985tv}. A crucial distinction of AdS black holes is their thermodynamic stability in certain regimes, enabled by a positive specific heat~\cite{Ashtekar:1999jx, Hawking:1982dh}. This allows them to be treated as well-defined thermal systems in equilibrium, supporting meaningful definitions of temperature, entropy, free energy, and thermodynamic potentials. These properties are essential for black hole thermodynamics and play a pivotal role in the AdS/CFT correspondence, where AdS black holes serve as powerful probes for investigating nonperturbative and strongly coupled phenomena in quantum field theories~\cite{Maldacena:1997re, Witten:1998qj}.
		The holographic description of strong-field regimes near black hole event horizons necessitates theories beyond linear Maxwell electrodynamics. This motivates (NED), which incorporates higher-order field corrections~\cite{Fradkin:1985qd,Tseytlin:1996it}. A foundational motivation for (NED) came from Born and Infeld~\cite{Born:1933lls, Born:1934gh}, who sought to eliminate the divergent self-energy of point charges by making the electric field finite at the origin.

        Nonlinear electrodynamics  serves dual purposes: it regularizes gravitational coordinate singularities and generates tunable holographic responses, making it ideal for studying strongly coupled systems via AdS/CFT~\cite{Ayon-Beato:1998hmi}. When coupled to gravity in AdS space, (NED)produces novel charged black hole solutions that differ significantly from standard Reissner-Nordström-AdS black holes~\cite{Ayon-Beato:1998hmi}. These solutions feature deformed asymptotic structures and modified horizon geometries~\cite{Yu:2019xdg}, while their nonlinear interactions substantially alter thermodynamic properties such as temperature and entropy.
		
		The singularity problem in black hole physics provides strong motivation for studying (NED) in the context of gravity. As established by Penrose~\cite{Penrose:1964wq}, gravitational collapse in General Relativity inevitably leads to curvature singularities, indicating a breakdown of classical gravity in deep interior regions and necessitating quantum or effective modifications. This raises the fundamental question of whether regular black holes—free of curvature singularities—can exist as exact solutions in modified gravity frameworks. The regular black hole metric satisfying the weak energy condition was constructed~\cite{Balart:2014jia}, although the underlying field equations were not specified. Later, the first exact electrically and magnetically charged regular black hole solutions were obtained through (NED) models minimally coupled to General Relativity~\cite{Bardeen:1965xi,Burinskii:2002pz}. For comprehensive reviews and current constraints on singularity resolution  via NED, see~\cite{Fan:2016hvf}.

        In references~\cite{Kubiznak:2012wp,Flores-Alfonso:2020euz}, the authors construct the first self-gravitating solutions within the unique, conformally invariant and duality-symmetric ModMax theory of nonlinear electrodynamics~\cite{Bandos:2020hgy,Bandos:2020jsw}, introducing a novel charged Taub-NUT spacetime and various black holes including Reissner-Nordström-type, accelerated, and AdS configurations. A key finding is that the nonlinear parameter acts as a screening factor, allowing for novel extremal black holes where the mass can be smaller than the charge~\cite{Barrientos:2022bzm}. Furthermore, the research develops a Melvin-Bonnor electromagnetic universe and related black holes embedded within it, revealing a novel Kerr-Schild representation for these spacetimes and expanding the spectrum of exact solutions to include vortex-like backgrounds~\cite{Barrientos:2024umq}. These results provide the first examples of accelerated and electromagnetized black holes in nonlinear electrodynamics, opening new avenues for studying strong-field regimes and raising fresh challenges for holographic interpretation~\cite{Barrientos:2024umq}.
        In~\cite{Bokulic:2025usc}, the authors  demonstrate that for a broad class of theories, regular black holes necessitate a mutual dependence between Komar mass and electric/magnetic charges, and prove that electrically charged black holes respecting the Maxwellian weak-field limit cannot have a bounded Kretschmann scalar. A key focus is the conformally invariant ModMax theory, for which they develop generalized Harrison transformations to generate new solutions, including multi-black hole configurations, black diholes, and their dilaton extensions~\cite{Bokulic:2025ucy}. These solutions often exhibit screening effects that alter the charge-to-mass ratio and can produce regular, extremal black holes without conical singularities. Collectively, this work underscores that while sophisticated solution-generating techniques can yield regular and multi-black hole spacetimes, achieving regularity through nonlinear electrodynamics often requires significant theoretical trade-offs~\cite{Bokulic:2025brf}. See more on ModMax black holes in~\cite{Bokulic:2021dtz,Pantig:2022gih,Kruglov:2021bhs,Kruglov:2022qag,Guzman-Herrera:2023zsv,Zhang:2021qga,EslamPanah:2024gxx,EslamPanah:2024qlu,Ahmed:2025qza,Heidari:2025llu,Canfora:2025gwm,Barrientos:2025rde}.
		
		Among various (NED)models, those constructed using the Courant--Hilbert representation~\cite{Friedlander_1963,Mkrtchyan:2022ulc,Russo:2024xnh} or derived from recent $T\bar{T}$-type deformations exhibit superior theoretical consistency, physical robustness, and often admit closed-form Lagrangians~\cite{Cavaglia:2016oda,Babaei-Aghbolagh:2022uij,He:2025ppz,Hou:2022csf,Smirnov:2016lqw,Bonelli:2018kik,Ferko:2022cix,Babaei-Aghbolagh:2024uqp}. Such models naturally respect electric–magnetic duality and  satisfy causality  conditions~\cite{Sorokin:2021tge,Bandos:2020jsw,Russo:2024llm,Russo:2024ptw}, and generate black hole solutions that are free from superluminal instabilities and acausal behavior near the event horizon~\cite{Chamblin:1999hg,Chamblin:1999tk,Kubiznak:2012wp,Cvetic:2010jb}. These features make them particularly attractive for gravitational  Ricci flows and holographic studies, where both classical coherence and quantum consistency are essential~\cite{Brizio:2024arr, Morone:2024sdg}.
		
In this framework, we focus on two recently proposed classes of nonlinear electrodynamic theories. One is generated in closed form by a $T\bar{T}$-deformation, while the other is a perturbative extension up to order $\lambda^3$, which reduces to the ModMax theory in the limit $\lambda \to 0$~\cite{Babaei-Aghbolagh:2025cni}. The GNED model preserves electric--magnetic duality and causality and provides a unified framework that includes, as special limits, various well-known nonlinear models such as ModMax, (GBI)~\cite{Bandos:2020hgy, Babaei-Aghbolagh:2022itg}, causal self-dual logarithmic electrodynamics~\cite{Russo:2024llm, Babaei-Aghbolagh:2025cni}, and $q$-deformed theories~\cite{Russo:2024ptw}. These properties make the GNED model a coherent and physically consistent framework for studying the responses of charged black holes and their holographic and thermodynamic behavior in asymptotically AdS spacetimes.
		In this study, we couple the general Lagrangian \( \mathcal{L}_{\mathrm{}}(F) \) to four-dimensional Einstein gravity with a negative cosmological constant \( \Lambda \leq 0 \). Our primary objective is to compute the electromagnetic potential and thermodynamic quantities of charged black holes and analyze their behavior under nonlinear electrodynamic effects. We focus on electrostatic configurations with spherical, planar, and hyperbolic horizon geometries, where the absence of magnetic fields simplifies the analysis while preserving the essential nonlinear characteristics of the theory.
		The corresponding bulk action takes the form
		\begin{equation}
			I = \frac{1}{16\pi} \int d^4 x \sqrt{-g} \left( R - 2\Lambda \right) + \frac{1}{4\pi} \int d^4 x \sqrt{-g} \, \mathcal{L}_{\mathrm{}}(S,P),
            \label{bulkActgned1}
		\end{equation}
		where  $ S=-\frac{1}{4}F_{\mu\nu}F^{\mu\nu}$ and $\,\,P=-\frac{1}{4}F_{\mu\nu}\tilde F^{\mu\nu}$ are two Lorentz invariant variables.
		
The black hole solution is derived by solving the generalized Einstein equations, sourced by the energy-momentum tensor of the (GNED) theory. This yields the metric function \( f(r) \), commonly referred to as the blackening function, which encodes the gravitational dynamics of the spacetime~\cite{Cai:2020wrp}. The event horizon is located at the radius \( r_h \) satisfying the condition \( f(r_h) = 0 \). 
Key thermodynamic quantities, including the Hawking temperature \( T \), entropy \( S \), and free energy \( F \), are determined from the near-horizon behavior of \( f(r) \) and its derivatives~\cite{Kubiznak:2012wp,Hendi:2012um,Cai:2014znn,Cai:2013qga,Xu:2015rfa}. Specifically, the Hawking temperature is given by:
\begin{equation}\label{Tempera}
T = \frac{1}{4\pi} \left. \frac{df(r)}{dr} \right|_{r = r_h},
\end{equation}
while the entropy follows the Bekenstein--Hawking area law \( S = \frac{A_h}{4} \), with \( A_h \) denoting the horizon area.

While Hawking radiation provides a semi-classical description of a black hole's exterior, the nature of its central singularity remains largely unknown~\cite{Hawking:1976ra,Hawking:1973uf,Hawking:1971vc,Townsend:1997ku}. The classical singularity theorems predict the existence of a singularity where general relativity breaks down, but its detailed character is not specified by these theorems~\cite{Szekeres:1960gm,Emparan:2008eg}. The asymptotic behavior of \( f(r) \) is analyzed both near the horizon and at spatial infinity to elucidate the geometric and physical characteristics of the spacetime~\cite{Joshi:2001xi,Goswami:2006ph}. Additionally, by computing curvature invariants such as the Kretschmann scalar \( \mathcal{K} =R_{\mu\nu\rho\sigma} R^{\mu\nu\rho\sigma} \), we assess the regularity of the solution and determine the presence or absence of curvature singularities~\cite{Virbhadra:2002ju,Virbhadra:2007kw,Gyulchev:2008ff,Sahu:2012er,Setare:2016dex,Adami:2017phg}.

The organization of this paper is as follows: In Section~\eqref{02}, we establish the theoretical foundation by introducing the Courant-Hilbert representation and the root-$T\bar{T}$ flow, which systematically generate a  (GNED) Lagrangian encompassing models such as ModMax, Born-Infeld, and logarithmic electrodynamics. In Section~\eqref{03}, we couple this (GNED) framework to gravity and derive novel black hole solutions, computing their key features including the metric function, electromagnetic potential, and horizon structure. Section~\eqref{04} presents the core thermodynamic analysis, where we calculate temperature, entropy, and free energy to reveal a rich phase structure exhibiting van der Waals-like transitions. In section~\eqref{055}, we also assess spacetime regularity by computing curvature invariants, demonstrating how (NED) interactions soften the central singularity. Finally, Section~\eqref{Discuss} concludes the work by discussing implications for the AdS/CFT correspondence and outlining prospects for future research on transport coefficients and fundamental bounds.

\section{ Causal Nonlinear Electrodynamics}\label{02} 

Nonlinear electrodynamics (NED) provides a consistent framework for extending Maxwell's theory beyond the linear regime while preserving fundamental principles such as gauge invariance, causality, and electric-magnetic duality~\cite{Bialynicki-Birula:1992rcm,Sorokin:2021tge,Gaillard:1981rj,Gaillard:1997rt,Gibbons:1995ap}. Recent developments demonstrate that causal, duality-invariant  (NED) models can be systematically constructed using the Courant-Hilbert (CH) formulation~\cite{Friedlander_1963}, which ensures the theory remains causal and satisfies convexity conditions~\cite{Russo:2024llm,Russo:2024xnh,Russo:2024ptw}.
In this formulation, the Lagrangian is derived from a single scalar generating function $\ell(\tau)$, where $\tau$ represents a specific combination of electromagnetic field invariants. This approach reduces the consistency conditions to straightforward algebraic or differential constraints on $\ell(\tau)$ and its derivatives. Remarkably, recent work has demonstrated that imposing electric-magnetic duality at the linearized level is sufficient to ensure complete causality in the nonlinear regime~\cite{Russo:2024llm}. Consequently, the $\mathrm{SO}(2)$ duality symmetry functions not merely as a physical requirement but as a fundamental structural principle governing the admissibility of nonlinear field dynamics.


The $\mathrm{SO}(2)$ electric--magnetic duality induces covariant rotations in the $(\mathcal{G}^{\mu\nu}, \tilde{F}^{\mu\nu})$ space, where the excitation tensor is defined as
\begin{equation}\label{Gmn}
	\mathcal{G}^{\mu\nu} \equiv -2 \frac{\partial \mathcal{L}}{\partial F_{\mu\nu}} = \mathcal{L}_S F^{\mu\nu} + \mathcal{L}_P \tilde{F}^{\mu\nu},
\end{equation}
where, $\mathcal{L}_S \equiv \frac{\partial \mathcal{L}}{\partial S}$ and $\mathcal{L}_P \equiv \frac{\partial \mathcal{L}}{\partial P}$. The $\mathrm{SO}(2)$ rotation acts linearly on the field doublet as
\begin{equation}
	\begin{pmatrix}
		\mathcal{G}'^{\mu\nu} \\
		\tilde{F}'^{\mu\nu}
	\end{pmatrix}
	=
	\begin{pmatrix}
		\cos \theta & \sin \theta \\
		-\sin \theta & \cos \theta
	\end{pmatrix}
	\begin{pmatrix}
		\mathcal{G}^{\mu\nu} \\
		\tilde{F}^{\mu\nu}
	\end{pmatrix},
\end{equation}
where the dual field strength is given by $\tilde{F}^{\mu\nu} = \frac{1}{2} \epsilon^{\mu\nu\lambda\kappa} F_{\lambda\kappa}$. This structure ensures that duality invariance is realized at the level of the equations of motion, provided the energy--momentum tensor is also invariant under such transformations.

Motivated by the SO(2) covariance of the field equations, several principal approaches have been developed to construct Lagrangians that respect electric--magnetic duality symmetry within NED. These methods ensure that the resulting equations of motion remain consistent with the underlying duality structure. Among these, the Gaillard-Zumino approach provides a systematic procedure for constructing invariant interaction terms that remain unchanged under duality transformations. The original work by Gaillard and Zumino established the groundwork, which was further elaborated by Gibbons and Rasheed~\cite{Gibbons:1995ap}. The Gaillard--Zumino--Gibbons--Rasheed framework requires (NED) Lagrangians $\mathcal{L}$ to satisfy the self-duality condition:
\begin{equation}
	\mathcal{G}\tilde{\mathcal{G}}-F\tilde{F}=0\,,
\end{equation}
as established in~\cite{Bialynicki-Birula:1992rcm,Gaillard:1981rj,Gaillard:1997rt,Gibbons:1995ap,Kuzenko:2000uh}.
The self-duality condition, in its differential form, can be expressed through the following partial differential equation (PDE):
\begin{eqnarray}
	\label{lagrangeNGZ}
	(\partial_S \mathcal{L})^2\, -2 \frac{S}{P} \, (\partial_S \mathcal{L}) (\partial_P \mathcal{L})-(\partial_P \mathcal{L})^2\, =1\, .
\end{eqnarray}

The condition (\ref{lagrangeNGZ}) ensures that the equations of motion derived from the Lagrangian maintain invariance under electric-magnetic duality transformations, a fundamental symmetry of (NED) \cite{Sorokin:2021tge}. To simplify the self-duality condition (\ref{lagrangeNGZ}), we introduce two non-negative variables:
\begin{eqnarray}
	U = \tfrac{1}{2}\left(\sqrt{S^2+P^2}-S\right), \label{UN} \qquad
	V = \tfrac{1}{2}\left(\sqrt{S^2+P^2}+S\right). \nonumber
\end{eqnarray}
In these variables, the self-duality equation takes the compact form \cite{Gaillard:1997rt}:
\begin{equation}
	\partial_U \mathcal{L} \, \partial_V \mathcal{L} = -1. \label{dUdN}
\end{equation}
The general solution to (\ref{dUdN}), derived through the Courant-Hilbert method \cite{Friedlander_1963}, is given by:
\begin{equation}
	\mathcal{L} = \ell(\tau) - \frac{2U}{\dot{\ell}(\tau)}, \qquad  \tau = V + \frac{U}{\dot{\ell}^2(\tau)} ,\label{lagg}
\end{equation}
where $\ell(\tau)$ denotes the Courant-Hilbert (CH) function with $\tau \geq 0$. The Maxwell theory corresponds to the simple case $\ell(\tau) = \tau$~\cite{Russo:2024llm}.
Remarkably, for weak-field (NED)  theories, the causality conditions reduce to simple constraints on the CH-function \cite{Russo:2024llm}:
\begin{equation}
	\dot{\ell} \geq 1, \quad \ddot{\ell} \geq 0, \label{CausalCondition}
\end{equation}
where the convexity requirement ($\ddot{\ell} \geq 0$) emerges naturally from the analysis. These universal conditions, independent of $(U, V)$, provide powerful constraints on viable self-dual (NED)  theories.
This flow is naturally formulated in the Courant–Hilbert (CH) representation, where the Lagrangian is governed by a generating function \( \ell(\tau) \), subject to the duality and causality bounds already introduced in Eq.(\ref{dUdN}). The energy-momentum tensor (EMT) associated with the Lagrangian ~\eqref{lagg} is
\begin{equation}\label{TMNRT}
	T_{\mu\nu} = \left(\frac{\tau\dot\ell}{U+V}\right) T^{\rm Max}_{\mu\nu} +
	(\ell - \tau \dot\ell) {\rm g}_{\mu\nu} \, ,  
\end{equation}
where $T^{Max}_{\mu\nu}=F_{\mu\rho}{F_{\nu}}^{\rho}-\frac{1}{4} \,g_{\mu\nu} F_{\alpha\beta}F^{\alpha\beta}$ represents the stress-energy tensor (EMT) of Maxwell's theory. Consequently, the trace of the EMT in any self-dual electrodynamic theory can generally be written as follows \cite{Russo:2024xnh}:
\begin{equation}\label{TnNl}
	{T_{\mu}}^\mu = 4(\ell - \tau \dot\ell)  \, .  
\end{equation}
The Courant--Hilbert (CH) representation provides a powerful framework for constructing causal and self-dual  (NED) models. Within this formalism, the theory is derived from a single generating function \(\ell(\tau)\), and the condition for a traceless energy-momentum tensor, \(\ell = \tau \dot{\ell}\), identifies a conformal fixed point invariant under marginal deformations. This structure is particularly well-suited for analyzing deformations driven by $\sqrt{T\bar{T}}$-type operators, as the trace depends solely on \(\ell(\tau)\)~\cite{Babaei-Aghbolagh:2025cni,Babaei-Aghbolagh:2025uoz}. Consequently, the CH representation unifies the classification of (NED)  theories and systematically describes their flow under such integrable deformations.
\section{Root $T\bar{T}$-Flow on Causal Nonlinear Electrodynamic
Theories}\label{03}
In the context of duality-invariant nonlinear electrodynamics, using Eq.~\eqref{TMNRT} enables the extraction of two fundamental scalar invariant structures from the energy-momentum tensor:
\begin{equation}\label{Tl}   
	{T_{\mu}}^{\mu} {T_{\nu}}^{\nu}= 16(\ell - \tau\dot\ell)^2\, ,   
	\qquad
	T_{\mu\nu}T^{\mu\nu}  =  4\left[(\tau\dot\ell)^2 + (\ell -\tau\dot\ell)^2\right] \, .  
\end{equation}
These invariants play a crucial role in characterizing the theory. The first structure —the square of the trace —vanishes for conformally invariant theories, while the second provides information about the nonlinear electromagnetic contributions to the energy density and pressure. These quantities are essential for constructing deformation operators, such as the irrelevant $T\bar{T}$ and  root-$T\bar{T}$ operators, and for analyzing the flow of thermodynamic and physical properties under such marginal deformations.

The irrelevant $T\bar{T}$ operator constitutes a specific composite function constructed from the energy-momentum tensor, defined as:
\begin{equation}
O_{\lambda}=\frac{1}{8} \left(T_{\mu\nu}T^{\mu\nu} - \frac{1}{2}{T_{\mu}}^{\mu} {T_{\nu}}^{\nu}\right).
\end{equation}
This operator represents a particular combination of the two independent quadratic invariants that can be formed from the energy-momentum tensor. Their difference, as featured in $O_{\lambda}$, isolates the traceless part of the energy-momentum tensor squared and serves as the defining structure for $T\bar{T}$-like deformations in various dimensions. This particular combination ensures that the operator maintains crucial properties, such as universality, solvability, and the preservation of symmetries under deformation.

The 4D root-$T\bar{T}$ deformation was first introduced in Ref.~\cite{Babaei-Aghbolagh:2022uij}, where this operator was identified within duality-invariant electromagnetic theories as describing the ModMax theory~\cite{Bandos:2020jsw}. This framework was subsequently extended to two-dimensional integrable models~\cite{Babaei-Aghbolagh:2025uoz,Babaei-Aghbolagh:2025hlm}. The root-$T\bar{T}$ operator in 4D is defined as:
\begin{equation}
\mathcal{R}_\gamma = \frac{1}{2}\sqrt{T_{\mu\nu}T^{\mu\nu} - \frac{1}{4}{T_{\mu}}^{\mu}{T_{\nu}}^{\nu}},
\end{equation}
and plays a crucial role in generating marginal deformations that preserve fundamental symmetries while introducing nonlinear interactions. This deformation has proven particularly valuable in constructing consistent, causal extensions of Maxwell electrodynamics and has opened new avenues for studying integrable structures in both higher-dimensional field theories and two-dimensional systems.
In Ref.~\cite{Babaei-Aghbolagh:2025cni}, a new representation for the root-$T\bar{T}$ operator is derived:
\begin{equation}
\mathcal{R}_\gamma = \tau\dot{\ell},
\end{equation}
which provides a unique relation for theories obtained from the Courant-Hilbert (CH) function in Eq.~\eqref{lagg}. The same work explores electrodynamic theories characterized by a marginal flow equation for the coupling $\gamma$. Based on the form of the marginal operator $\mathcal{R}_\gamma$, we propose that all causal theories must satisfy the flow equation:
\begin{equation}\label{Rootsc}
	\frac{\partial \mathcal{L}}{\partial \gamma} = \mathcal{R}_\gamma = \tau\dot{\ell}.
\end{equation}
This equation establishes a fundamental connection between the deformation of the Lagrangian with respect to the coupling parameter $\gamma$ and the geometric structure encoded in the CH formalism.
In the following sections, we proceed to systematically review and analyze the aforementioned theories, including ModMax, Generalized Born-Infeld (GBI), self-dual logarithmic electrodynamics, and causal $q$-deformed electrodynamics models, with particular focus on their structural properties under root-type marginal deformations.
\subsection{ ModMax theory }\label{0.1} 

Maxwell electrodynamics is, of course, both conformal and duality-invariant. However, most non-linear generalizations, such as the well-known Born-Infeld theory, break conformal invariance. The ModMax theory fills this gap by being the simplest non-linear, one-parameter deformation that maintains these fundamental symmetries. The ModMax theory, proposed by Bandos, Lechner, Sorokin, and Townsend~\cite{Bandos:2020jsw}, represents a significant development in theoretical electrodynamics. It is the \emph{unique} non-linear, gauge-invariant extension of Maxwell's theory in four spacetime dimensions that preserves both conformal invariance and the SO(2) electric--magnetic duality invariance
\begin{equation}\label{ModM}
	\mathcal{L}_{\text{ModMax}} = S \cosh\gamma + \sinh\gamma\, \sqrt{S^2 + P^2}.
\end{equation}
Here, $\gamma$ is a dimensionless deformation parameter that governs the strength of nonlinearity. For physical consistency, such as causality, unitarity, and energy condition requirements, it is necessary to impose $\gamma \geq 0$, which ensures that the Lagrangian remains convex in the electric field.

In the CH formalism, the CH function for this theory must be linear in $\tau$, and we therefore take $\ell(\tau) = e^{\gamma} \tau$, scaled by the coupling constant $\gamma$. This choice implies $\dot{\ell} = e^{\gamma}$, and it follows that $\ell = \tau \dot{\ell}$. As shown in \cite{Russo:2024xnh}, the condition $\ell = \tau \dot{\ell}$ establishes the traceless condition for ModMax theories. This CH function yields:

\begin{equation}
	\label{LMM}
	\mathcal{L}_{MM} = e^\gamma V - e^{-\gamma} U \,, \qquad  \qquad  \tau= e^{-2 \gamma}\,U  + V. 
\end{equation}
Using Eq.~\eqref{UN}, we can express the Lagrangian~\eqref{LMM} in terms of the variables $S$ and $P$ as~\eqref{ModM}.
This form automatically satisfies the self-duality constraint~\eqref{dUdN} and respects the causality and convexity conditions given by Eq.~\eqref{CausalCondition}.

According to Eq.~\eqref{TnNl}, in ModMax theory we have \( T^{\mu}_{\ \mu} = 0 \), which confirms the exact preservation of conformal symmetry within the Courant-Hilbert (CH) framework. Moreover, the theory satisfies the root-type \( T\bar{T} \) trace flow condition given by Eq.~\eqref{Rootsc}; in other words, it acts as a conformal fixed point of the root-\( T\bar{T} \) flow and remains dynamically stable under marginal integrable deformations that preserve causality 
and duality. This implies that ModMax not only exhibits \( T\bar{T} \)-like flows~\cite{Babaei-Aghbolagh:2022uij, Conti:2018jho},  but also represents a consistent and stable boundary theory from the perspective of effective field theory, compatible with fundamental 
symmetries~\cite{Bandos:2020hgy, Russo:2024xnh}.
\subsection{ Generalized Born-Infeld theory}\label{0.3} 
As shown in Ref.~\cite{Babaei-Aghbolagh:2022uij}, the (GBI) theory can be obtained by applying a root-type \( T\bar{T} \) deformation to the ModMax theory~\cite{Bandos:2020hgy}. The resulting theory describes a Born--Infeld-type nonlinear electrodynamics that preserves both causality and duality symmetry. It admits a CH representation with a generating function given by
\begin{equation}
	\label{llMMB}
	\ell(\tau) = \frac{1}{\lambda} - \sqrt{\frac{1}{\lambda} \left( \frac{1}{\lambda} - 2e^{\gamma} \tau \right)}\,.
\end{equation}
Considering~\eqref{llMMB}, solving second equation~\eqref{lagg} yields the following expression for $\tau$:
\begin{equation}
	\tau = \frac{U + e^{2\gamma} V}{e^{\gamma} (2\lambda U + e^{\gamma})}.
\end{equation}
The corresponding Lagrangian density derived from the first equation~\eqref{lagg} reads
\begin{equation}
	\label{MMB}
	\mathcal{L}_{GBI} = \frac{1}{\lambda} - \sqrt{ \left( \frac{1}{\lambda} + 2e^{-\gamma} U \right) \left( \frac{1}{\lambda} - 2e^{\gamma} V \right) }\,,
\end{equation}
which remains real and well-defined under the causality condition~\eqref{CausalCondition}.
An alternative expression of the Lagrangian~\eqref{MMB} in terms of the scalar invariants \( S \) and \( P \) is given by:
\begin{equation}
	\label{LEGBI}
	\mathcal{L}_{GBI} = \frac{1}{\lambda} \left( 1 - \sqrt{ 1 - 2\lambda \left( S\cosh\gamma + \sqrt{S^2 + P^2} \sinh\gamma \right) - \lambda^2 P^2 } \right).
\end{equation}
It can be explicitly demonstrated that the (GBI) theory satisfies the self-duality condition in~\eqref{lagrangeNGZ}. Furthermore, as shown in~\cite{Russo:2024xnh}, the causality condition given in ~\eqref{CausalCondition} is satisfied by the Lagrangian defined in ~\eqref{LEGBI}.
Recently, Nastase~\cite{Nastase:2021ffq} investigated the possibility of interpreting (GBI) electrodynamics within a brane-like construction in string theory. This particular formulation reduces to the Modified Maxwell (ModMax) theory in the weak-field limit. The origin of both ModMax theory and its (GBI) has been elucidated in~\cite{Babaei-Aghbolagh:2022uij}, where they were shown to emerge from a $T\bar{T}$-like deformation of Maxwell's theory.
\subsection{Causal  Self-Dual Logarithmic Electrodynamics }
\label{2.3}
The duality-invariant theory of ``logarithmic electrodynamics,'' which incorporates the couplings $\lambda$ and $\gamma$, can be reduced to the ModMax theory in the weak-field limit and is known to adhere to the causality principle \cite{Russo:2024llm}.
For this logarithmic electrodynamics, we define the CH function as\footnote{This CH-function corresponds to the one introduced in \cite{Russo:2024llm} for the specific parameter value $\lambda = (e^{\gamma} T)^{-1}$.}:
\begin{equation}\label{tauLog}
	\ell (\tau)=- \frac{1}{\lambda} \log(1 -  e^{\gamma} \lambda \tau )\,.
\end{equation} 
The expansion of the logarithmic CH function is:
\begin{eqnarray}\label{ExplLog1}
	\ell (\tau) =e^\gamma \,  \tau + \frac{1}{2} \, \lambda \,  e^{2\gamma} \, \tau^2 +\frac{1}{3}\, \lambda^2 \,  e^{3\gamma} \, \tau^3+\frac{1}{4}\, \lambda^3 \,  e^{4\gamma} \, \tau^4  +\cdots .
\end{eqnarray}
By solving the second equation~\eqref{lagg}, for  
CH-function~\eqref{tauLog}, the auxiliary field $\tau$ can be determined as follows:
\begin{equation}\label{tau}
	\tau= \frac{1}{2  \lambda^2 U} + \frac{1}{e^{\gamma} \lambda}- \frac{\sqrt{1 + 4  \lambda U (e^{- \gamma} -   \lambda V)}}{2  \lambda^2 U}.
\end{equation}
By substituting the value of $\tau$ from equation~\eqref{tau} into the Lagrangian~\eqref{lagg}, we derive a logarithmic electrodynamic theory in the form of:
\begin{equation}\label{Log}
	\mathcal{L}_{Log} =\frac{1}{\lambda} \biggl( 1 -  \sqrt{1 + 4  \lambda U (e^{- \gamma} -   \lambda V)} -  \log\biggl(\frac{e^{\gamma} \Bigl(-1 + \sqrt{1 + 4  \lambda U (e^{- \gamma} -   \lambda V)}\Bigr)}{2  \lambda U}\biggr) \biggr)\,.
\end{equation}
This logarithmic Lagrangian satisfies the PDE duality condition~\eqref{dUdN} and adheres to the principle of causality, fulfilling condition~\eqref{CausalCondition} provided that $\gamma \geq 0$.
Within the CH-representation framework, the logarithmic electrodynamics theory exhibits a non-vanishing trace of the energy-momentum tensor. This signals an explicit breaking of conformal symmetry within the model. Nevertheless, the theory satisfies the root-type deformation flow condition given in Eq.~\eqref{Rootsc} and consequently remains stable under integrable deformations of the $T\bar{T}$ type. As a result, logarithmic electrodynamics constitutes a causal and self-dual theory.

\subsection{ Causal Self-Dual  No Maximum-$\tau$ Electrodynamics }
\label{2.4}
A particularly interesting example of a self-dual and causal nonlinear electrodynamics model is obtained by choosing a CH-function with a fractional power structure~\cite{Russo:2024llm,Russo:2024xnh}:
\begin{equation}\label{CHRoot}
	\ell(\tau) = -\frac{2}{3\lambda}\left(1 - \left(1 + e^\gamma \lambda \tau\right)^{3/2}\right).
\end{equation}
The derivative of this CH-function is:
\begin{equation}\label{dotlRoot}
	\dot\ell(\tau) = e^\gamma \sqrt{1 + e^\gamma \lambda \tau}. 
\end{equation}
Substituting into the general formalism, the resulting Lagrangian for this theory reads:
\begin{equation}\label{RootLag}
	\mathcal{L} = \frac{2\sqrt{2}}{3\lambda} \left(1 + e^\gamma \lambda V - \frac{\Delta}{2}\right) \sqrt{1 + e^\gamma \lambda V + \Delta} - \frac{2}{3\lambda},
\end{equation}
where the \(\Delta\) is defined as:
\begin{equation}\label{DeltaRoot}
	\Delta = \sqrt{(1 + e^\gamma \lambda V)^2 + 4 \lambda e^{-\gamma} U}.
\end{equation}
This model is constructed within the Courant--Hilbert (CH) representation using a generating function that satisfies both the duality condition~\eqref{dUdN} and the causality condition~\eqref{CausalCondition}, provided that $\gamma \geq 0$. The theory exhibits scale-breaking behavior at finite $\lambda$, as indicated by the non-vanishing trace of its energy--momentum tensor. However, in the limit $\lambda \to 0$, it smoothly reduces to the conformal ModMax theory, thereby asymptotically restoring conformal symmetry.

\subsection{Causal q-deform Electrodynamics }
\label{2.5}

Finally, causal \(q\)-deformed Electrodynamics also belongs to the class of models satisfying condition~\eqref{Rootsc}, confirming the consistency of the CH-based formalism and its dependence on the nonlinearity parameter \(\gamma\).
Within the framework of CH representations, a broad class of duality-invariant and causal \(q\)-deformed nonlinear electrodynamics theories can be systematically constructed by choosing CH-functions with specific fractional powers~\cite{Babaei-Aghbolagh:2025uoz}. The corresponding CH-function takes the following form~\cite{Russo:2024ptw,Russo:2024xnh}:
\begin{eqnarray}\label{qDef}
	\ell (\tau) =\frac{1}{\lambda} \Big(1 -  (1 - \tfrac{1}{\mathbf{q}} e^{\gamma} \lambda \tau )^{\mathbf{q}}\Big)\,. 
\end{eqnarray}
This structure is constrained by the initial condition \(\ell(\lambda = 0, \tau) =e^{\gamma} \tau\), which ensures consistency with the ModMax limit.
In the $q$-deformed, self-dual (electromagnetic $\mathrm{U}(1)$-duality invariant) nonlinear electrodynamic theories, causality is respected under the conditions $\dot{\ell} \ge 1$ and $\ddot{\ell} \ge 0$. For the $q$-deformed theory, these conditions lead to the parameter range $\frac{1}{2} \leqslant q \leqslant 1$.
The expansion of the $q$-deformed Courant–Hilbert (CH) function around $\lambda = 0$ is given by
\begin{equation}\label{qD}
    \ell(\tau) = e^{\gamma} \tau - \frac{e^{2\gamma}(q - 1)}{2q} \lambda \tau^2 + \frac{e^{3\gamma}(q - 2)(q - 1)}{6q^2} \lambda^2 \tau^3 - \frac{e^{4\gamma}(q - 3)(q - 2)(q - 1)}{24q^3} \lambda^3 \tau^4 + \cdots\,. \nonumber
\end{equation}
For the specific values $q = \tfrac{3}{2}$ and $q = \tfrac{3}{4}$, the parameter $\tau$ can be determined analytically by solving the second equation~\eqref{lagg} in terms of the auxiliary variables $U$ and $V$.
A dimensional reduction of these theories from four to two dimensions produces a family of integrable nonlinear models, continuously parametrized by a  $q$ parameter~\cite{Babaei-Aghbolagh:2025uoz,Babaei-Aghbolagh:2025hlm}.
\subsection{General Causal Theories }\label{222.11} 

In this section, we examine a class of (NED) theories that reduce to the ModMax model in the non-interacting limit via a $T\bar{T}$ deformation. For all causal models, the associated Courant–Hilbert (CH) function $\ell(\tau)$ admits a power series expansion around $\tau = 0$ governed by a dimensionful coupling constant $\lambda$:
\begin{eqnarray}\label{expan}
    \ell(\tau) = e^\gamma \tau + \lambda\, \mathcal{O}(\tau^2)\,.
\end{eqnarray}
Theories characterized by CH functions satisfying ~\eqref{expan} and parametrized by $(\gamma, \lambda)$ consistently flow to ModMax as $\lambda \to 0$. As shown in~\cite{Babaei-Aghbolagh:2025cni}, any such expansion ensures compatibility with the root flow equation~\eqref{Rootsc}, even when extended to include arbitrary higher-order terms. To construct these (GNED) models, we systematically expand the CH function as follows:
\begin{equation}\label{Lexpan}
	\ell (\tau) =e^\gamma \,  \tau  +\sum_{i=1}^\infty m_i \, \lambda^i\, e^{(i+1)\gamma} \, \tau^{i+1} ,
\end{equation}
where $m_i$ are constants.
The CH-function~\eqref{Lexpan} defines a broad class of nonlinear  (NED) theories that are both causal and electric–magnetic duality-invariant when expanded perturbatively in powers of $\lambda$. Each coefficient $m_i$ in the expansion encodes the contribution at order $\lambda^i$, enabling systematic classification and generation of models~\cite{Babaei-Aghbolagh:2025cni, Babaei-Aghbolagh:2024uqp}.
Given an appropriate set of $m_i$ coefficients, one can construct the CH-function to any desired order in $\lambda$. All such theories, characterized by the two couplings $(\gamma, \lambda)$ and satisfying the structure of~\eqref{Lexpan}, reduce to the ModMax theory in the limit $\lambda \to 0$.
By expanding the CH-function~\eqref{Lexpan} up to $\mathcal{O}(\lambda^3)$ and applying the second equation in~\eqref{lagg}, the auxiliary field $\tau$ can be computed to third order as follows:
\begin{eqnarray}\label{Gq1q22}
	\tau&=&e^{-2 \gamma} U + V - 4\,  m_1 \, \lambda e^{-3 \gamma} U (U + e^{2 \gamma} V)+ \lambda^2 e^{-4 \gamma} \bigl( 4 m_1^2 U (U + e^{2 \gamma} V) (7 U + 3 e^{2 \gamma} V)\nonumber\\
	&&-6 m_2 U (U + e^{2 \gamma} V)^2 \bigr)+4 \lambda^3 e^{-5 \gamma} U \Bigl(-2 m_3 (U + e^{2 \gamma} V)^3 -  m_1 (U + e^{2 \gamma} V) \nonumber\\
	&&\times\bigl(-9 m_2 (U + e^{2 \gamma} V) (3 U + e^{2 \gamma} V)+ 4 m_1^2 (5 U + e^{2 \gamma} V) (3 U + 2 e^{2 \gamma} V)\bigr)\Bigr) .  
\end{eqnarray}
Using the first equation in~\eqref{lagg} together with the $\tau$ expansion~\eqref{Gq1q22}, we derive the effective Lagrangian in terms of the $\tau$ field variables $U$ and $V$. By expressing $U$ and $V$ in terms of the standard electromagnetic invariants $S$ and $P$ via the relation~\eqref{UN}, we obtain the full nonlinear Lagrangian $\mathcal{L}_{\text{GNED}}$ expanded up to order $\lambda^3$.
\begin{eqnarray}\label{Gq1q2n11}
	\mathcal{L}_{GNED} &=& S \cosh(\gamma) + \sqrt{P^2 + S^2} \sinh(\gamma) + m_1 \lambda \bigl(\sqrt{P^2 + S^2} \cosh(\gamma) + S \sinh(\gamma)\bigr)^2 \\
	&&+\lambda^2 \Bigl(-2 e^{-\gamma} m_1^2 \bigl(- S + \sqrt{P^2 + S^2}\bigr) \bigl(\sqrt{P^2 + S^2} \cosh(\gamma) + S \sinh(\gamma)\bigr)^2 \nonumber\\
	&&+ m_2 \bigl(\sqrt{P^2 + S^2} \cosh(\gamma) + S \sinh(\gamma)\bigr)^3\Bigr) +\lambda^3 \biggl(\tfrac{1}{2} e^{-4 \gamma} m_1^3 \Bigl((1 + e^{2 \gamma})^2 (3 + e^{2 \gamma}) P^4\nonumber\\
	&& + 24 S^3 \bigl(S - \sqrt{P^2 + S^2}\bigr) + 2 P^2 S \bigl((12 + 7 e^{2 \gamma} + e^{6 \gamma}) S + (-6 - 7 e^{2 \gamma} + e^{6 \gamma})\sqrt{P^2 + S^2}\bigr)\Bigr) \nonumber\\
	&&-  \frac{e^{-4 \gamma}}{4 }3 m_1 m_2 \Bigl((1 + e^{2 \gamma})^3 P^4 + 8 S^3 \bigl(S - \sqrt{P^2 + S^2}\bigr)\nonumber\\
	&& + 2 (1 + e^{2 \gamma}) P^2 S \bigl(4 S + (-2 + e^{2 \gamma}) (1 + e^{2 \gamma})\sqrt{P^2 + S^2} + 2 e^{3 \gamma} S \sinh(\gamma)\bigr)\Bigr) + \nonumber\\
	&&\frac{1}{8} m_3 \Bigl(3 P^4 + 4 P^2 (P^2 + 2 S^2) \cosh(2 \gamma) + (P^4 + 8 P^2 S^2 + 8 S^4) \cosh(4 \gamma) \nonumber\\
	&&+ 8 S\sqrt{P^2 + S^2} \bigl(P^2 + (P^2 + 2 S^2) \cosh(2 \gamma)\bigr) \sinh(2 \gamma)\Bigr)\biggr)\nonumber\,.  
\end{eqnarray}
This expression \eqref{Gq1q2n11} encodes the general form of a causal nonlinear electrodynamics theory that is both perturbatively consistent and duality-symmetric up to third order in $\lambda$. Each term in the expansion reflects an increasing contribution from the nonlinear self-interactions of the electromagnetic field. For different choices of the coefficients $m_1, m_2, m_3$, one can recover specific known models as particular limits of this general framework.
All causal theories up to any order of $\lambda$ can be obtained from the CH-function expansion \eqref{Lexpan}. Given a set of coefficients $\{m_i\}$ appearing in the general Lagrangian \eqref{Gq1q2n11}, we can systematically derive a perturbative expansion of the Lagrangian of any causal theory to arbitrary order in the coupling constant $\lambda$. All the theories studied in Refs.~\cite{Russo:2024xnh} are summarized in Table~\eqref{T11}, showing that the extensions of these theories form a special subgroup of the general Lagrangian \eqref{Gq1q2n11} when an appropriate set of $m_i$ coefficients is chosen.
\begin{table}[H] 
	\caption[]{\label{T11} A general causal theory vs. All causal theories }
	\centering
	\begin{tabular}{|c|c|c|c|}\hline
		Models & $\lambda^1$-Order & $\lambda^2$-Order & $\lambda^3$-Order    \\ \hline
		GBI   theory &  $m_1=\frac{1}{2}$& $m_2=\frac{1}{2}$& $m_3=\frac{5}{8}$  \\  \hline
		Logarithmic (NED)  theory & $m_1=\frac{1}{2}$& $m_2=\frac{1}{3}$& $m_3=\frac{1}{4}$  \\ \hline
		No maximum-$\tau$ (NED)  theory & $m_1=\frac{1}{4}$& $m_2=-\frac{1}{24}$& $m_3=\frac{1}{64}$  \\ \hline
		$q=2/3$-deform (NED)  theory &  $m_1=\frac{1}{4}$& $m_2=\frac{1}{6}$& $m_3=\frac{7}{48}$ \\ \hline
		$q=3/4$-deform (NED)  theory &  $m_1=\frac{1}{6}$& $m_2=\frac{5}{54}$& $m_3=\frac{5}{72}$   \\ \hline
	\end{tabular}
	\label{tab1}
\end{table}

In the context of gravitational physics, coupling a general and causal  (GNED) Lagrangian to Einstein gravity yields a diverse range of black hole solutions in asymptotically AdS spacetimes. The Lagrangian, derived from a CH expansion in powers of the deformation parameter $\lambda$, systematically encodes higher-order nonlinear electromagnetic interactions. These interactions significantly modify the spacetime geometry near the black hole horizon, where linear Maxwell theory fails to provide an accurate description.
By tuning the expansion coefficients $\{m_1, m_2, m_3, \dots\}$ in Eq.~\eqref{Gq1q2n11}, one can recover well-known models such as the Born–Infeld, logarithmic, and $q$-deformed theories as special cases. This unified framework enables a systematic study of thermodynamic quantities, including entropy, temperature, specific heat, and free energy, as well as critical behavior and phase transitions in black hole systems.

Thus, the (GNED) Lagrangian serves as a unifying tool for generating and classifying a broad class of black hole solutions while maintaining fundamental consistency conditions such as causality and electric–magnetic duality. Moreover, the nonlinear and duality-covariant structure of (GNED) provides a physically motivated basis for constructing consistent holographic duals of strongly coupled quantum matter. In particular, it governs the transport properties of the dual theory, such as electrical conductivity and heat diffusion, through the nonlinear structure of the electromagnetic sector.

\section{Gravity Setup and  Thermodynamics of Black Holes  }\label{04}
Our understanding of matter is inextricably linked to our comprehension of spacetime itself. This connection was first rigorously formulated by Einstein, whose theory of general relativity describes gravity as a manifestation of spacetime curvature induced by mass and energy. Einstein's field equations elegantly relate the geometry of spacetime (encoded in the Einstein tensor $G_{\mu\nu}$) to its energy-momentum content (described by $T_{\mu\nu}$). A profound prediction of these equations is the existence of black holes: regions of spacetime bounded by an event horizon, from which no classical information can escape.

Studies of (NED) black holes reveal consistent thermodynamic behavior across different frameworks and horizon topologies. The analysis in \cite{Chemissany:2008fy} of Nonlinear Electrodynamic (NED) black holes in Einstein-Born-Infeld gravity establishes a consistency between the entropy function formalism and direct entropy calculations. As reviewed in \cite{Liang:2019dni} and \cite{Bai:2022vmx}, these systems exhibit rich phase structure: thermal AdS space can be the globally stable phase, with a first-order, Hawking-Page-like transition to BI-AdS black holes at small electric potential that becomes zeroth order at larger potential values. Furthermore, works in \cite{Cai:1996eg,Cai:2004eh,Wang:2019kxp,Yang:2022qoc}, shows that while BI-AdS black holes with flat or hyperbolic horizons are always thermodynamically stable, their spherical counterparts have a critical Born-Infeld parameter, above which they are always stable and below which an unstable phase appears.
\subsection{$AdS$ black holes for General-Born-Infeld Theory}\label{041}
The framework for our analysis is a four-dimensional theory of gravity described by the Einstein-GBI action, which is further generalized by the inclusion of a cosmological constant, $\Lambda$. This constant is essential for modeling a universe with a non-zero vacuum energy.
\begin{equation}
			I = \frac{1}{16\pi} \int d^4 x \sqrt{-g} \left( R - 2\Lambda \right) + \frac{1}{4\pi} \int d^4 x \sqrt{-g} \, \mathcal{L}_{\mathrm{GBI}}(F).
            \label{GraBI}
		\end{equation}
The first term presents the standard Einstein-Hilbert action, featuring the Ricci scalar $R$, while the second term describes the action for the (GBI) theory of electrodynamics.
The four-dimensional Einstein–Hilbert theory coupled to the general nonlinear causal electrodynamics is described by the bulk action given in Eq.~\eqref{bulkActgned1}.  
Varying the action ~\eqref{GraBI} with respect to the metric $g_{\mu\nu}$ yields the Einstein field equations:
\begin{equation}\label{EinE}
	R_{\mu\nu} - \frac{1}{2} R\, g_{\mu\nu} + \Lambda g_{\mu\nu} = 2 \, T_{\mu\nu} \, ,
\end{equation}
where $T_{\mu\nu}$ is the energy–momentum tensor of the GBI  nonlinear electrodynamics sector.

Varying the full action with respect to the fundamental fields yields the system's equations of motion. Specifically, variation with respect to the gauge potential 
$A_\mu$ gives the (GBI) electromagnetic field equations, while variation with respect to the metric 
$g_{\mu \nu}$ produces the Einstein equations with a cosmological constant.
Using definition~\eqref{Gmn}, we can find the antisymmetric tensor $\mathcal{G}^{\mu\nu}$ for the (GBI) Lagrangian \eqref{LEGBI} as follows:
\begin{eqnarray}
			\mathcal{G}_{\mu\nu} &=&\frac{\cosh(\gamma) \sqrt{P^2 + S^2} + S \sinh(\gamma)}{\sqrt{ P^2 + S^2} \sqrt{1 + \lambda^2 P^2 - 2 \lambda \bigl(\cosh(\gamma) S + \sqrt{P^2 + S^2} \sinh(\gamma)\bigr)}} F_{\mu\nu}\\ &+& \frac{P \bigl(- \lambda \sqrt{P^2 + S^2} + \sinh(\gamma)\bigr)}{\sqrt{P^2 + S^2} \sqrt{1 + \lambda^2 P^2 - 2 \lambda \bigl(\cosh(\gamma) S + \sqrt{P^2 + S^2} \sinh(\gamma)\bigr)}} \tilde{F}_{\mu\nu}.\nonumber
            \label{Gmunu}
\end{eqnarray}

In nonlinear electrodynamics, the electromagnetic field is described using the displacement field $\mathbf{D}$ and the magnetic field $\mathbf{H}$, which are general functions of the electric field $\mathbf{E}$ and the magnetic flux density $\mathbf{B}$. The fields $\mathbf{E}$ and $\mathbf{B}$ are the fundamental dynamical variables generated by physical sources and governed by Maxwell's equations.
In nonlinear media, the constitutive relations defining $\mathbf{D}$ and $\mathbf{H}$ are given by:
$
\mathbf{H} = \mathbf{H}(\mathbf{E}, \mathbf{B}), \quad \mathbf{D} = \mathbf{D}(\mathbf{E}, \mathbf{B}).
$
A common specific form for these relations, often derived from a Lagrangian density $\mathcal{L}$, is:
\begin{eqnarray}\label{HDE}
\mathbf{H} = -\frac{\partial \mathcal{L}}{\partial \mathbf{B}}, \quad \mathbf{D} = \frac{\partial \mathcal{L}}{\partial \mathbf{E}}.
\end{eqnarray}
The gauge potential is assumed to take the electrostatic form $A_\mu = (A_t(r), 0, 0, 0)$. With these definitions, the following result is obtained for the (GBI) theory:
\begin{eqnarray}
			\mathcal{G}_{0\, 1} =D_r=- \frac{e^{\gamma} A_t^{\prime}(r)}{\sqrt{1 -  \lambda  \, e^{\gamma} {A_t^{\prime}(r)}^2}}=\frac{Q}{r^2}.
            \label{Gtx}
\end{eqnarray}
From the solution of~\eqref{Gtx}, we find the time component of the electric potential $A_t(r)$ as
\begin{eqnarray}\label{AtGBI}
		A_t(r) = \frac{e^{-\gamma} Q}{r} \, {}_2F_1(\tfrac{1}{4}, \tfrac{1}{2}, \tfrac{5}{4}, -\lambda \frac{e^{-\gamma} Q^2}{ r^4}).
\end{eqnarray}
Consider an $AdS$ background with radius $L$ and cosmological constant $\Lambda = -3/L^2$. The metric ansatz is then
\begin{equation}\label{HigherDMetric}
	ds^2 = -f(r)dt^2 + \frac{dr^2}{f(r)} + r^2 d\Omega^2_{k},
\end{equation}
where the AdS curvature scale is denoted by $L$, and the parameter $k \in \{-1,0,+1\}$ specifies the spatial curvature of the two-dimensional metric $d\Omega^2_{k}$.  
The three possible constant-curvature geometries, corresponding to hyperbolic, planar, and spherical cases, are explicitly given by
\begin{equation}\label{geometries}
	d\Omega^2_{k}=\left\lbrace\begin{matrix}
		& d\theta^2+\sin^2\theta\, d\phi^2_{}\ \ \ &\ \  {\rm for\ }k=+1\ \,,\cr
		& \,\,\,\ dx^2+d y^2\ \ \qquad
		\qquad&\ \ {\rm for\ }k=0\ \ \ \,,\cr
		&d\theta^2+\sinh^2\theta\, d\phi^2_{}&\ \  {\rm for\ }k=-1\,.
	\end{matrix}\right. 
\end{equation}
Substituting the energy--momentum tensor of the GBI theory and the metric ansatz~\eqref{HigherDMetric} into Einstein's field equations~\eqref{EinE} yields the following differential equations:
\begin{eqnarray}\label{EinT}
	f(r) + \frac{1}{\lambda}f^{\prime}(r) \lambda r + ( \lambda \,\Lambda-2 ) r^2 + 2 \sqrt{e^{-\gamma} Q^2 \lambda + r^4}&=&0\\
  \tfrac{1}{2} r \bigl(2 f^{\prime}(r) + (f^{\prime \prime}(r) + 2 \Lambda) r\bigr) - 2 \bigl(\frac{r^2}{\lambda} -  \frac{r^4}{\lambda \sqrt{e^{-\gamma} Q^2 \lambda + r^4}}\bigr)  &=&0\nonumber
\end{eqnarray}
Solving the differential equations~\eqref{EinT} leads to the following blackening factor for the $AdS$ metric \eqref{HigherDMetric}:
\begin{eqnarray}\label{frGBI}
		f(r) = k- \frac{m}{r} + \frac{2 r^2}{3 \lambda} -  \frac{2 \sqrt{ \lambda\, e^{-\gamma} \, Q^2+r^4 }}{3 \lambda} -  \tfrac{1}{3} r^2 \Lambda + \frac{4 e^{-\gamma} Q^2 }{3 r^2} {}_2F_1(\tfrac{1}{4}, \tfrac{1}{2}, \tfrac{5}{4}, - \tfrac{e^{-\gamma} Q^2 \lambda}{r^4}),
\end{eqnarray}
where $m$ is the mass parameter.
The horizon radius $r_h$, defined by the largest root of $f(r_h)=0$, is related to the mass parameter $m$ through
\begin{eqnarray}\label{massGBI}
		m = k \,r_h  -  \tfrac{1}{3} r_h^3 \Lambda + \frac{2 r_h }{3 \lambda} \bigl(r_h^2 -  \sqrt{\lambda e^{-\gamma} Q^2 + r_h^4}\bigr)+ \frac{4 e^{-\gamma} Q^2 }{3 r_h} \,{}_2F_1(\tfrac{1}{4}, \tfrac{1}{2}, \tfrac{5}{4}, - \tfrac{\lambda e^{-\gamma} Q^2}{r_h^4}).
\end{eqnarray}
The corresponding black hole mass is then given by
\begin{equation}\label{MassBI}
	M = 2\Omega_{k}  m.
\end{equation}
The dimensionless volume $\Omega_{k}$ takes different forms depending on the spatial geometry.
\begin{figure}[h]
	\begin{subfigure}{0.45\textwidth}\includegraphics[width=\textwidth]{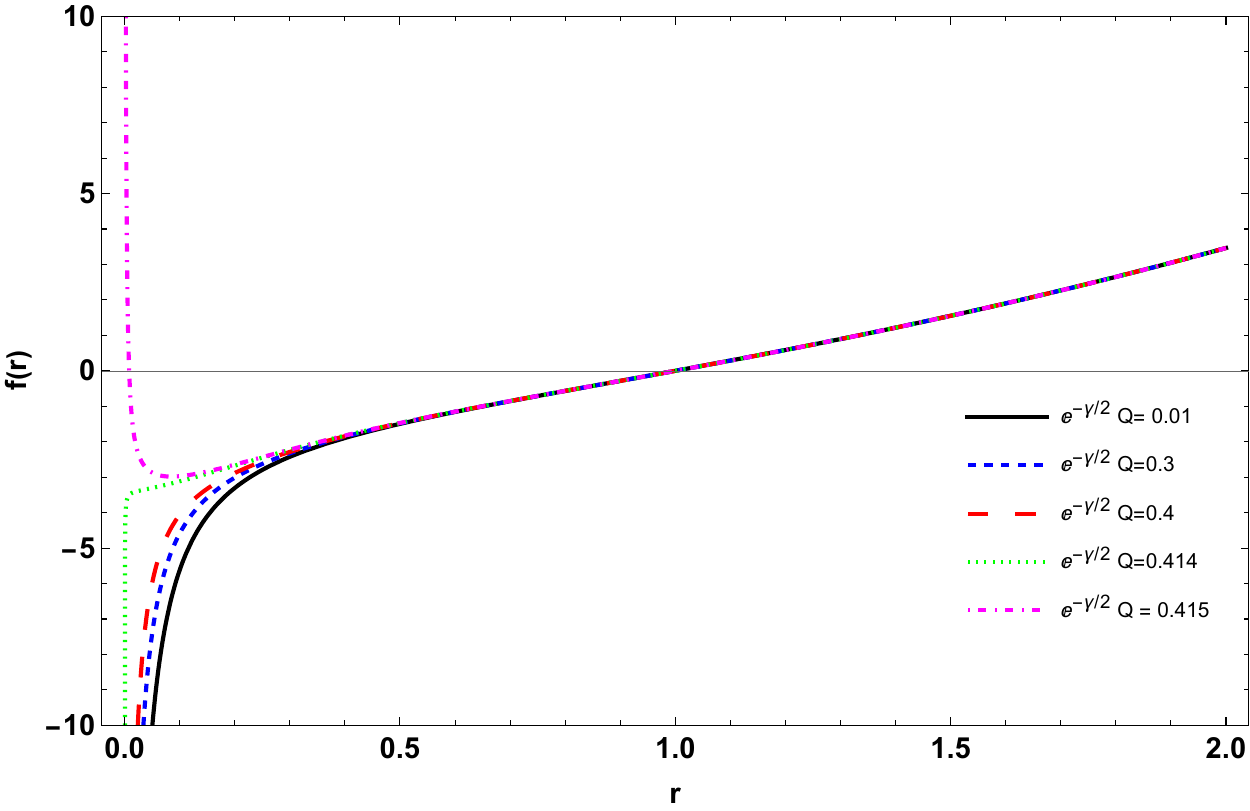}
		\caption{}
		\label{fig:frBIQ}
	\end{subfigure}
 \begin{subfigure}{0.45\textwidth}\includegraphics[width=\textwidth]{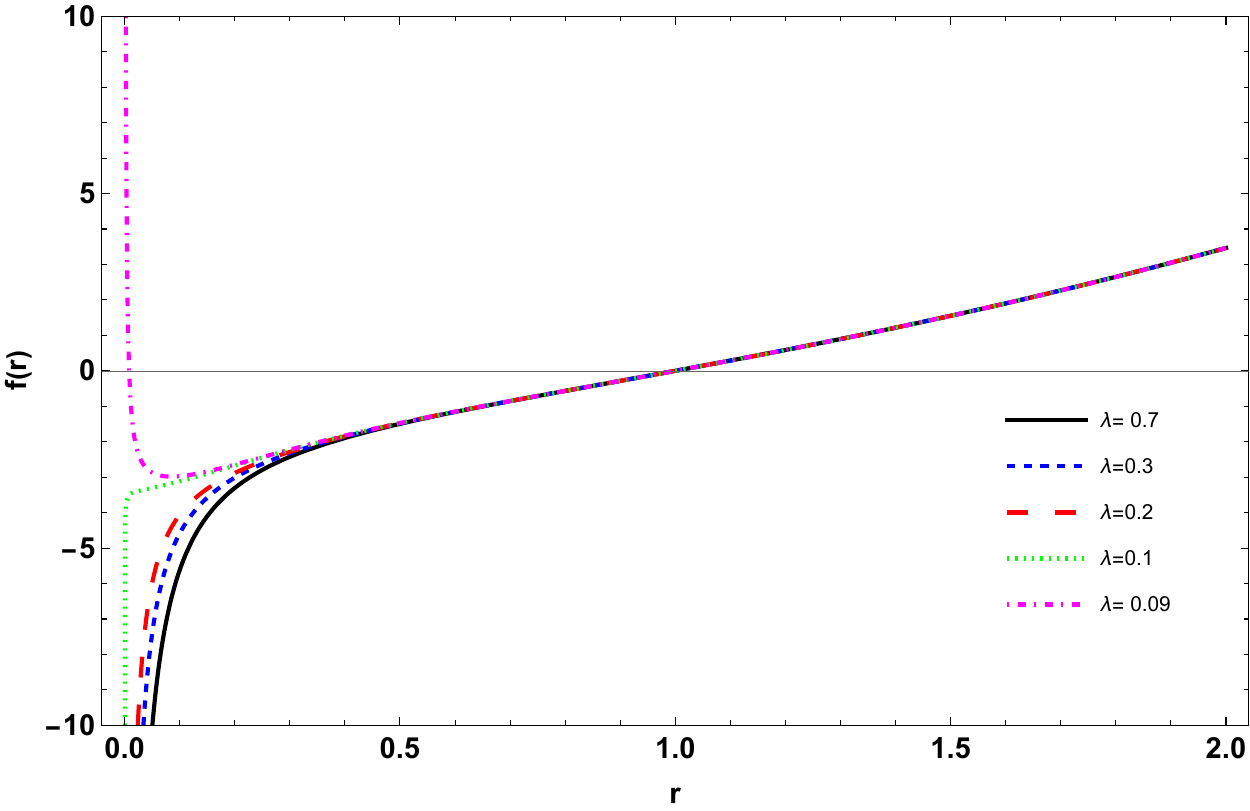}
		\caption{}
		\label{fig:frBIlam}
	\end{subfigure}
	\caption{ Plots of the metric function $f(r)$ versus $r$ for Born-Infeld gravity with $k=0$, $L=1$, and $r_h = 1$ are shown. In Fig.~\eqref{fig:frBIQ}, with $\lambda = 0.1$, the black hole transitions from a single-horizon to a two-horizon state as the electric charge $Q$ and $\gamma$ are increased beyond a threshold of $e^{-\gamma/2}Q\approx 0.414$. Conversely, in Fig.~\eqref{fig:frBIlam}, with a fixed charge and $\gamma$ at $e^{-\gamma/2}Q = 0.410$, the same transition to a two-horizon state occurs as the Born-Infeld parameter $\lambda$ is decreased below a value of $\lambda \approx 0.1$. }
	\label{fig:1rhBI}
\end{figure}

In Fig.~\eqref{fig:1rhBI}, we analyze the horizon structure for the case $k=0$, where the black hole may exhibit either a single horizon or two distinct horizons, depending on the values of the parameters. A similar qualitative behavior is observed for $k=1$: within a certain parameter range, the solution admits a single-horizon configuration, which subsequently transitions to a two-horizon phase. In the presence of two horizons, the inner horizon corresponds to the Cauchy horizon.
A qualitatively different behavior arises for $k=-1$. In this case, the two-horizon configuration is always present, while a single-horizon solution never occurs. Consequently, a Cauchy horizon is unavoidable for $k=-1$. This feature is of particular importance, as it highlights a fundamental distinction between the singularity structure and the interior geometry of black holes arising in nonlinear electrodynamics (NED).  The interior geometry of the black hole is highly sensitive to variations in the effective electric charge $(\tilde{Q}=e^{-\gamma/2}Q)$ and coupling parameters $(\lambda)$, indicating that the internal behavior is governed by the competition between the gravitational mass and the self-energy.

As shown in Ref.~\cite{Hale:2025ezt,Hale:2025urg},  when the gravitational mass dominates over the self-energy, the interior geometry resembles that of a Schwarzschild black hole, without Cauchy horizon. In contrast, when the self-energy exceeds the mass, the spacetime develops a Reissner-Nordström–like structure, characterized by the presence of a Cauchy horizon and a timelike singularity.
A general form of the black factor $f(r)$ for all electrodynamic theories coupled to gravity can be expressed as 
\begin{eqnarray}\label{fgencf}
f(r) = k- \frac{m}{r}+ \frac{U(r)}{r}\,,
\end{eqnarray}
where \( U(r) \) is the self-energy density of the electromagnetic theory, obtained from the stress-energy tensor \(U(r) =- \frac{1}{2 \pi}  \int T^0\,_{0} \, dV \). For any theory, the self-energy density can be expanded near $r=0$ in the form 
\begin{eqnarray}\label{expfff}
U(r) =\frac{U^{(-1)}}{r}+U^{(0)}+r U^{(1)}+\mathcal{O}(r^2)\,.
\end{eqnarray} 
As shown in~\cite{Hale:2025ezt}, for any theory that, like Maxwell's theory, has a nonzero (and positive) inverse function \(U^{(-1)} \), Cauchy horizons will appear. In the (GBI) case, the expansion of the metric function \( f(r) \) for large \( r=0 \) is given by:
\begin{eqnarray}\label{fGBIEx}
f(r) = k- \frac{m}{r} +\frac{ e^{-\tfrac{3}{4}\gamma} Q^{3/2} \Gamma(\tfrac{1}{4})^2}{r\,(3 \pi^{1/2} \lambda^{1/4})} - \frac{2 e^{-\tfrac{\gamma}{2} } Q}{\lambda^{1/2}}+\mathcal{O}(r)\,.
\end{eqnarray}
Since in this theory \( U^{(-1)} = 0 \) and the model does not admit a Cauchy horizon. We know \( U^{(0)} \) denote the self-energy of a point charge located at \( r = 0 \). In this model, this quantity satisfies
\begin{eqnarray}\label{SelfEBI}
 U^{(0)} =\frac{ e^{-\tfrac{3}{4}\gamma} Q^{3/2} \Gamma(\tfrac{1}{4})^2}{3 \pi^{1/2} \lambda^{1/4}} ,
\end{eqnarray}
which is finite. Consequently, the self-energy of a charged particle in the theory is regular.\footnote{We would like to thank Robie A. Hennigar and Jorge G. Russo for discussions on this point.
}
In Sec.~\eqref{055}, we investigate the internal structure of this class of black holes and determine the exact charge-to-mass ratio that separates causal black holes from singular configurations.
\subsubsection{Thermodynamics of GBI-AdS black holes}\label{4111}
The thermodynamic quantities of charged AdS black holes in GBI theory are characterized by the Hawking temperature~\eqref{Tempera} and the Bekenstein–Hawking entropy, $S = 4 \pi \,\Omega_{k}\, r_h^{2}$.
We can obtain the temperature by using Eqs.~\eqref{Tempera} and~\eqref{frGBI} as
\begin{eqnarray}\label{TTGBI}
		T = \frac{1}{4\pi} \Bigg(\frac{k}{r_h} + \frac{2 r_h }{\lambda}-  \frac{2 }{\lambda \,r_h} \sqrt{\lambda e^{-\gamma}  Q^2 +r_h^4  } -  r_h \Lambda  \Bigg).
\end{eqnarray}
By defining the electric charge density,
\begin{equation}
    \rho= \frac{Q}{4\pi}\Omega_k
\end{equation}
it is straightforward to show that the thermodynamic quantities, mass~\eqref{MassBI}, temperature~\eqref{TTGBI}, and entropy, satisfy the first law of black hole thermodynamics:
\begin{equation}
    dM = T\,dS + \Phi\,d\rho,
\end{equation}
where $\Phi$ is the electrostatic potential at the black hole horizon, conjugate to the electric charge density $\rho$.
Using Eq.~\eqref{AtGBI}, the electrostatic potential difference $\Phi$ between the horizon and infinity can be expressed as
\begin{equation}
  \Phi=16 \pi  \frac{e^{-\tfrac{\gamma}{2}} \, Q}{r_h}  \, {}_2F_1(\tfrac{1}{4}, \tfrac{1}{2}, \tfrac{5}{4}, -\lambda \frac{e^{-\gamma} Q^2}{ r_h^4}).
\end{equation}

The thermodynamic behavior and phase structure of charged AdS black holes can be 
systematically analyzed using the Helmholtz free energy, which serves as a central 
quantity for identifying possible phase transitions~\cite{Chamblin:1999hg,Chamblin:1999tk,Kubiznak:2012wp,Cvetic:2010jb}.
In the canonical ensemble with fixed electric charge $Q$, the free energy is defined as
\begin{equation}\label{F}
    F(T) = M - T S,
\end{equation}
where $M$ is the mass, $T$ is the Hawking temperature and $S$ is the Bekenstein-Hawking entropy.
In the present GBI model, the corresponding Helmholtz free energy 
takes the form
\begin{eqnarray}\label{freeGBI}
		F = \Omega_{k} \bigl(k\, r_h + \frac{2 r_h}{3 \lambda} \sqrt{r_h^4 +\lambda e^{-\gamma} Q^2 }+ \frac{r_h^3 \Lambda}{3 } -\frac{2 \, r_h^3 }{3 \lambda}+ \frac{8 e^{-\gamma} Q^2 }{3 r_h} {}_2F_1(\tfrac{1}{4}, \tfrac{1}{2}, \tfrac{5}{4}, - \tfrac{\lambda e^{-\gamma} Q^2 }{r_h^4}) \bigr).\nonumber\\
\end{eqnarray}

\begin{figure}[h]
	\begin{subfigure}{0.45\textwidth}\includegraphics[width=\textwidth]{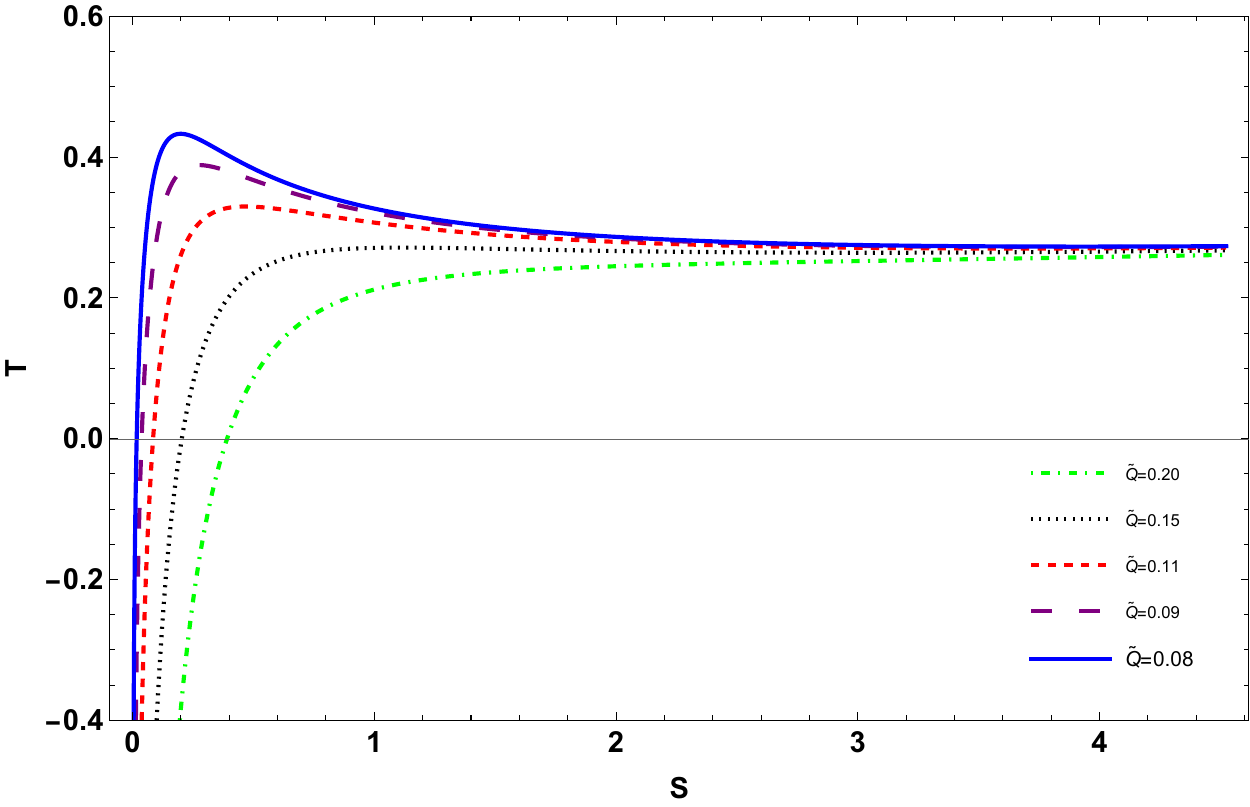}
		\caption{}
		\label{fig:TSBIQ}
	\end{subfigure}
 \begin{subfigure}{0.45\textwidth}\includegraphics[width=\textwidth]{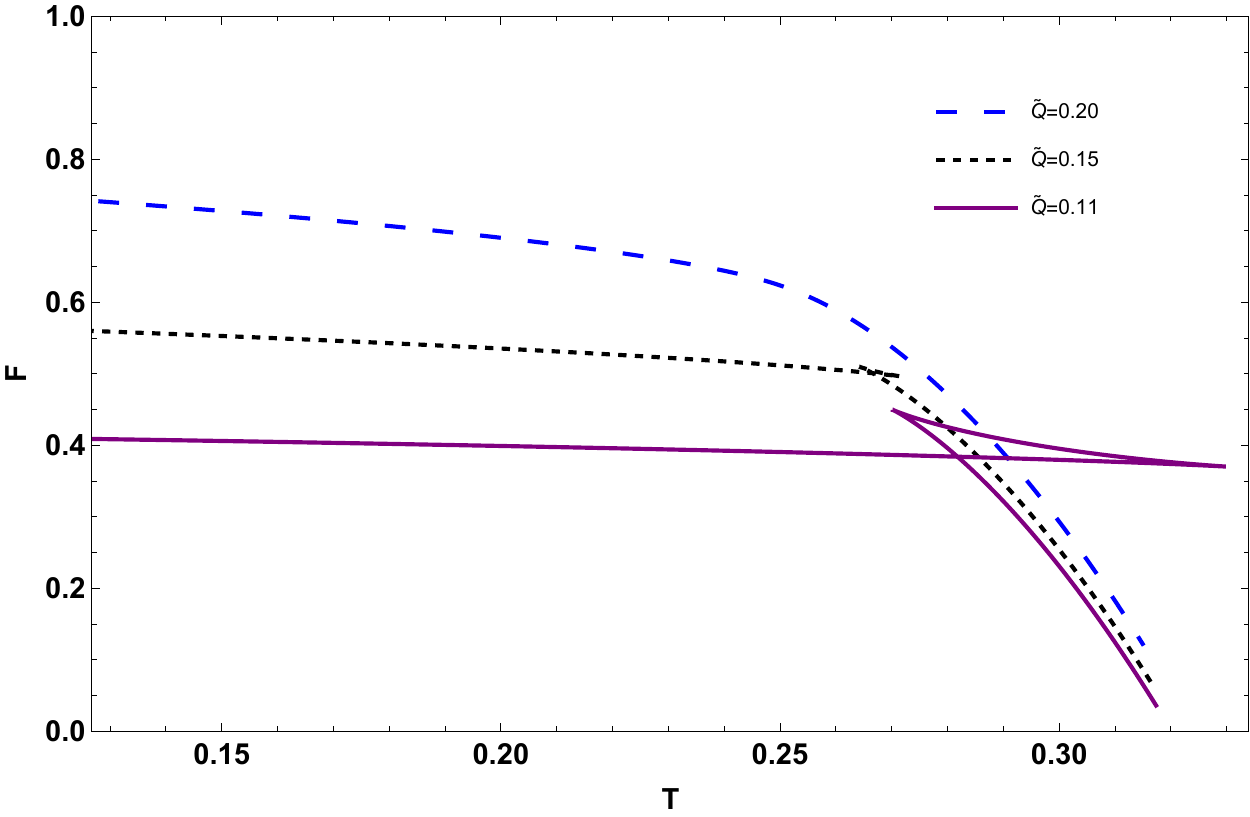}
		\caption{}
		\label{fig:FTBIQ}
	\end{subfigure}
	\caption{ \ref{fig:TSBIQ}: The temperature vs entropy  for (GBI) black hole for spherical ($k=1 ,L=1$) case by ﬁxed $\lambda=0.02$ for different variable effective electric charge $\tilde{Q}=e^{-\gamma/2}Q$, and 
    \ref{fig:FTBIQ}: The Helmholtz free energy vs temperature for a (GBI) black hole for the spherical ($k=1 ,L=1$) case, ﬁxed in  $\lambda=0.02$ for different variable effective electric charge $\tilde{Q}=e^{-\gamma/2}Q$. }
	\label{fig:FTBI}
\end{figure}
In Fig.~\eqref{fig:FTBI}, we see that a GBI-AdS black hole exhibits thermodynamic behavior similar to that of a van der Waals gas.
Fig.~\eqref{fig:TSBIQ} displays the temperature ($T$)  as a function of entropy ($S$)
for the GBI–AdS black hole with spherical topology ($k=1$). We plot the phase transition diagrams in the $T$--$S$ plane in Fig.~\ref{fig:TSBIQ}. From Fig.~\ref{fig:TSBIQ}, it can be observed that for the charged GBI-AdS black hole in the ensemble of fixed charge, a Van der Waals-type phase transition appears in the $T$--$S$ plane. This behavior is completely analogous to that of the Van der Waals liquid-gas system in general.
Fig.~\eqref{fig:FTBIQ} displays the Helmholtz free energy ($F$) as a function of temperature ($T$)
for the GBI–AdS black hole with spherical topology ($k=1$). As shown in Fig.~\eqref{fig:FTBIQ}, the $F$--$T$ diagram typically exhibits a characteristic ``swallowtail'' structure, indicating a first-order phase transition between the small and large black hole branches. This transition corresponds to a discontinuity in the first derivative of the free energy with respect to temperature.
At the coexistence temperature $T_c$, the two branches intersect, satisfying
\begin{equation}
F_{\mathrm{small}}(T_c) = F_{\mathrm{large}}(T_c),
\end{equation}
which implies that both phases can coexist in thermal equilibrium. For $T < T_c$, the small black hole phase minimizes $F(T)$ and is thermodynamically preferred, whereas for $T > T_c$, the large black hole phase dominates.
As the system approaches the critical charge $Q_c$, the swallowtail structure gradually disappears. At the critical point, the second derivative $(\partial^2 F / \partial T^2)_Q$ vanishes, marking a continuous (second-order) phase transition. This behavior is analogous to the liquid--gas critical point in the Van der Waals fluid.
\subsection{$AdS$ black holes for  Logarithmic Theory}\label{059}
A general discussion of logarithmic black holes can be found in \cite{Soleng:1995kn,Kruglov:2019ybs,Ali:2020jeb,Ali:2023zgm}, where it is shown that such theories do not satisfy the causality principle in the presence of a magnetic field.
In this section, we analyze black hole solutions in logarithmic nonlinear electrodynamics~\cite{Russo:2024llm,Russo:2024xnh}, defined by the CH function~\eqref{tauLog}. This model, a causal and duality-preserving transformation of Maxwell's theory, has recently been investigated in the causal Courant-Hilbert framework~\cite{Russo:2024xnh}, which clarifies its connection to $T\bar{T}$-like deformations~\cite{Babaei-Aghbolagh:2025cni}.
A refined analysis of causal and duality-symmetric (NED) theory  was carried out by Russo and Townsend~\cite{Russo:2024xnh}, 
who analyzed the logarithmic model in the general CH representation, where $\gamma$ denotes the deformation parameter controlling the logarithmic branch of the theory~\eqref{Log}. 
For an electrostatic configuration with charge \(Q\) and gauge field \(A_\mu = (A_t(r),\,0,\,0,\,0)\), the magnetic field vanishes.
Under this condition, the Lorentz variable \(P\) is defined to be zero. This definition implies that the \(U\) vanishes and \(V = S\). Consequently, the Lagrangian of the logarithmic theory can be written in the following form:
\begin{equation}\label{Lagsssss}
		\mathcal{L}_{Log} \;=\;- \frac{1}{\lambda} \log\bigl( 1 -\lambda   e^{\gamma} S \bigr).
\end{equation}
The first step in finding solutions to Einstein's equations for a given background is to compute the electric potential and determine the corresponding energy-momentum tensor, \( T_{\mu\nu} \), which is coupled to gravity within the logarithmic (NED)  model.
In this framework, we first extract the tensor \( \mathcal{G}_{\mu\nu} \) from the Lagrangian density \( \mathcal{L}_{Log} \) as follows:
\begin{equation}\label{RussoGmunulog}
		\mathcal{G}_{\mu\nu} \;=\;- \frac{e^{\gamma} F_{\mu \nu }}{ \lambda  e^{\gamma} S-1}.
\end{equation}
Considering an electrostatic configuration with charge \( Q \) and a gauge potential of the form \( A_\mu = (A_t(r), 0, 0, 0) \), and substituting this ansatz into Eq.~\eqref{RussoGmunulog}, we obtain:
\begin{equation}\label{RussoSingularityglog}
	\mathcal{G}_{01} \;=\;D_r=\, \frac{2 e^{\gamma} A_t^{\prime}(r)}{\lambda \,  e^{\gamma}  {A_t^{\prime}(r)}^2-2 }\,=\,\frac{Q}{r^2} .
\end{equation}
By solving the equation above, the electromagnetic potential $A_t(r)$ can be determined as a function of the radial coordinate $r$, expressed in terms of the electric charge $Q$ and the coupling constants $\lambda$ and $\gamma$ as follows:
\begin{eqnarray}\label{ATrlog}
	A_t(r) &=&\frac{r^3}{3 Q \lambda} -  \frac{r \sqrt{r^4 + 2 e^{-\gamma} Q^2 \lambda}}{3 Q \lambda} + \frac{4 e^{-\gamma} Q \,}{3 r}{}_2F_1(\tfrac{1}{4}, \tfrac{1}{2}, \tfrac{5}{4}, - \tfrac{2 e^{-\gamma} Q^2 \lambda}{r^4}).
\end{eqnarray}
The energy-momentum tensor, \( T_{\mu\nu} \), for the logarithmic electrodynamic theory is derived from the Lagrangian density, \( \mathcal{L} \), given in Eq.~\eqref{Lagsssss} via the relation:
\begin{eqnarray}\label{TTlogg}
T_{\mu \nu} &=& \frac{e^{\gamma} F_{\mu }{}^{\alpha } F_{\nu \alpha }}{1 -  e^{\gamma} \lambda S} -  \frac{\log(1 -  e^{\gamma} \lambda S) \mathit{g}_{\mu \nu}}{\lambda}.
\end{eqnarray}
By substituting the electric potential from Eq.~\eqref{ATrlog} and the metric from Eq.~\eqref{HigherDMetric} with \( k = 0 \) into the definition of the energy-momentum tensor, we can compute its non-zero components. Consequently, for the logarithmic electrodynamic theory, the energy-momentum tensor \( T^{\mu}_{\phantom{\mu}\nu} \) is determined to be:
\begin{eqnarray}\label{TQQLog}
T_{\mu \nu}=\left(\begin{array}{cccc}
\frac{\biggl(- r^2 + \log\bigl(X\bigr) r^2 + \sqrt{\frac{2 Q^2 \lambda}{e^{\gamma}} + \
r^4}\biggr) \mathcal{F} (r)}{\lambda r^2} & 0 & 0 & 0\\
0 & \frac{r^2 - \log\bigl(X\bigr) r^2 - \
\sqrt{\frac{2 Q^2 \lambda}{e^{\gamma}} + r^4}}{\lambda r^2 \
\mathcal{F} (r)} & 0 & 0\\
0 & 0 & - \frac{\log\bigl(X\bigr) r^2}{\
\lambda} & 0\\
0 & 0 & 0 & - \frac{\log\bigl(X\bigr) r^2}{\lambda}
\end{array}\right),\nonumber
\end{eqnarray}
where $\log\bigl(X\bigr)=\log\bigl(\frac{e^{\gamma} \bigl(- r^4 + r^2 \sqrt{\frac{2 \
Q^2 \lambda}{e^{\gamma}} + r^4}\bigr)}{Q^2 \lambda}\bigr).$

\begin{figure}[h]
	\begin{subfigure}{0.45\textwidth}\includegraphics[width=\textwidth]{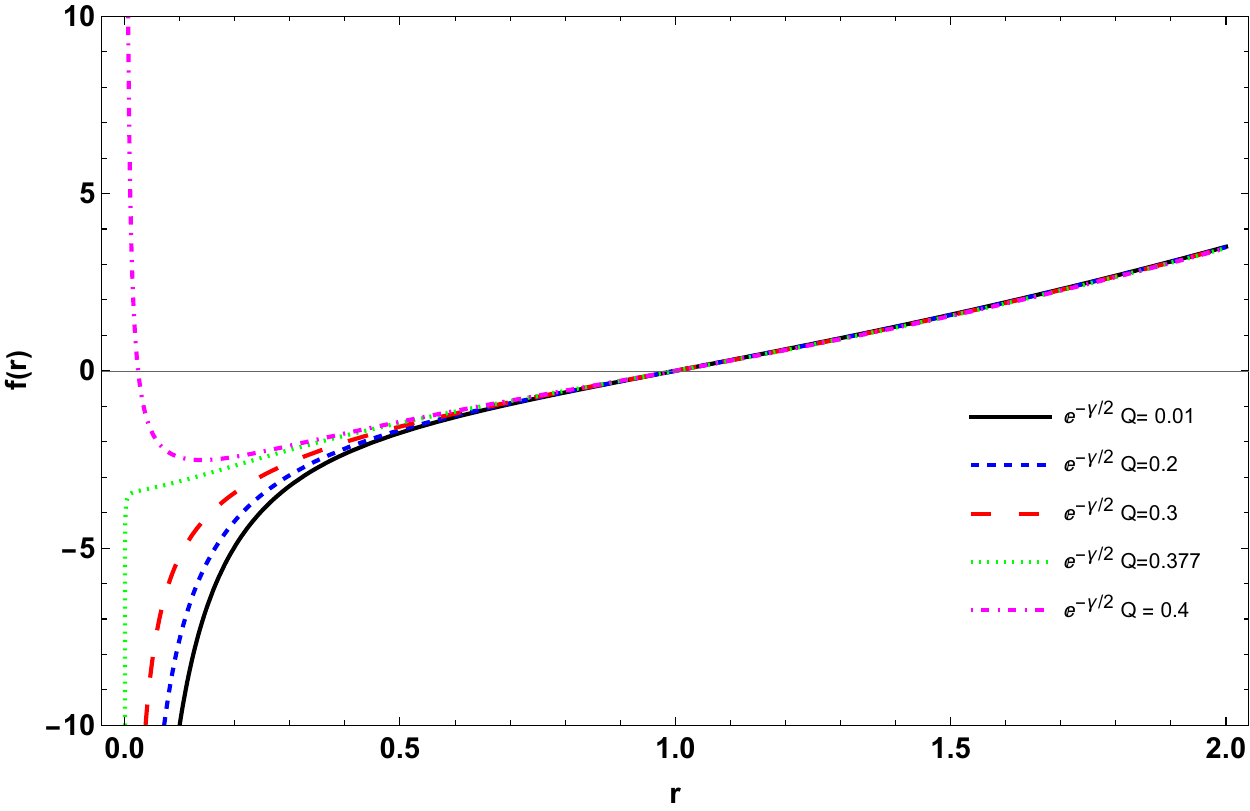}
		\caption{}
		\label{fig:frLogQ}
	\end{subfigure}
 \begin{subfigure}{0.45\textwidth}\includegraphics[width=\textwidth]{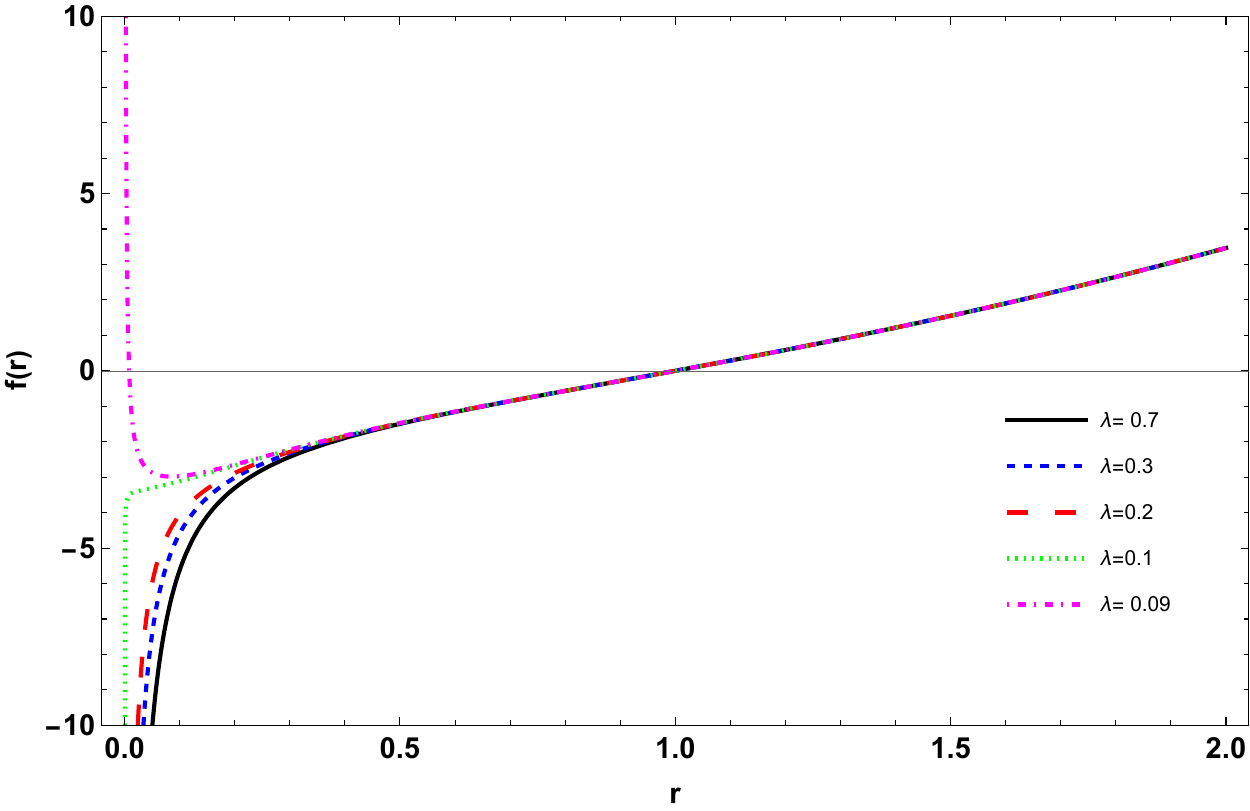}
		\caption{}
		\label{fig:frLoglam}
	\end{subfigure}+
	\caption{ Plots of the metric function $f(r)$ versus $r$ for Log-$AdS$ gravity with $k=0$, $L=1$, and $r_h = 1$ are shown. In Fig.~\eqref{fig:frLogQ}, with $\lambda = 0.1$, the black hole transitions from a single-horizon to a two-horizon state as the electric charge $Q$ and $\gamma$ is increased beyond a threshold of $e^{-\gamma/2}Q\approx 0.377$. Conversely, in Fig.~\eqref{fig:frLoglam}, with a fixed charge and $\gamma$ at $e^{-\gamma/2}Q = 0.377$, the same transition to a two-horizon state occurs as the coupling  $\lambda$ is decreased below a  value of $\lambda \approx 0.1$. }
	\label{fig:Logrh}
\end{figure}

As a result, in the general case where \( k \neq 0 \), the Einstein field equations for the logarithmic theory, given by Eq.~\eqref{EinE}, take the following form:
\begin{eqnarray}\label{Eq1}
&&f(r) + f^{\prime}(r)\, r-k -  \frac{2 r^2}{\lambda} + r^2 \Lambda + \frac{2 \sqrt{r^4 + 2 e^{-\gamma} Q^2 \lambda}}{\lambda} \\
&&+ \frac{2 r^2 }{\lambda} \log\bigl( \frac{ r^2 \sqrt{r^4 + 2 e^{-\gamma} Q^2 \lambda}-r^4  }{e^{-\gamma} Q^2 \lambda}\bigr)=0,\nonumber\\
&&2 f^{\prime}(r)+ r \Bigl(f^{\prime \prime}(r) + 2 \Lambda + \frac{4 }{\lambda} \log\bigl( \frac{r^2\sqrt{r^4 + 2 e^{-\gamma} Q^2 \lambda}-r^4 }{e^{-\gamma} Q^2 \lambda}\bigr)\Bigr)=0.\nonumber
\end{eqnarray}
An analytical solution to the Einstein field equations can be obtained starting from the framework established in Eq.~\eqref{Eq1}. The following line element give the corresponding metric
\begin{eqnarray}\label{fr}
 f(r)& =&k- \frac{m}{r} + \frac{10 r^2}{9 \lambda} -  \tfrac{1}{3} r^2 \Lambda -  \frac{10  \sqrt{r^4 + 2 \lambda e^{-\gamma} Q^2 }}{9 \lambda} \nonumber+\frac{16 e^{-\gamma} Q^2 }{9 r^2} {}_2F_1(\tfrac{1}{4}, \tfrac{1}{2}, \tfrac{5}{4}, - \tfrac{2 e^{-\gamma} Q^2 \lambda}{r^4})\\
 &-&  \frac{2 r^2 }{3 \lambda} \log\Bigl(\frac{r^2 \bigl(  \sqrt{r^4 + 2 \lambda e^{-\gamma} Q^2 }- r^2 \bigr)}{ \lambda e^{-\gamma} Q^2}\Bigr).
\end{eqnarray}
The mass parameter \( m \) is determined by solving the horizon condition \( f(r_h) = 0 \), which yields the following expression in terms of the black hole's horizon radius \( r_h \):
\begin{eqnarray}\label{masslog}
m& =&k \,r_h+\frac{10 r_h^3}{9 \lambda}  -  \tfrac{1}{3} r_h^3 \Lambda-  \frac{10 r_h \sqrt{r_h^4 + 2 e^{-\gamma} Q^2 \lambda}}{9 \lambda}\nonumber+\frac{16 e^{-\gamma} Q^2 }{9 r_h} {}_2F_1(\tfrac{1}{4}, \tfrac{1}{2}, \tfrac{5}{4}, - \tfrac{2 e^{-\gamma} Q^2 \lambda}{r_h^4})\\
&-&  \frac{2 r_h^3 }{3 \lambda} \log\Bigl(\frac{r_h^2 \bigl( \sqrt{r_h^4 + 2 e^{-\gamma} Q^2 \lambda}\bigr)- r_h^2 }{e^{-\gamma} Q^2 \lambda}\Bigr).
\end{eqnarray}
The corresponding black hole mass $M$ is then given by:
\begin{equation}\label{Masslogg}
	M = 2\Omega_{k}  m.
\end{equation}
The horizon structure for gravity with Logarithmic nonlinear electrodynamics theory in the case of a flat geometry (\( k = 0 \)) is shown in Fig.~\eqref{fig:Logrh}.  Specifically, we find that the black hole can exhibit two distinct configurations based on its parameters: a case featuring a single, degenerate horizon, and  another case possessing two distinct horizons—an outer event horizon and an inner Cauchy horizon. This behavior  is regulated by the model's specific coupling constant \( \lambda \).

In the Logarithmic nonlinear electrodynamics black hole, the expansion of the metric function \( f(r) \) for large \( r=0 \) in \eqref{fr} is given by:
\begin{eqnarray}\label{flojEx}
f(r) = k- \frac{m}{r} +\frac{2 \times 2^{\tfrac{3}{4}} e^{-\tfrac{3}{4}\gamma} Q^{3/2} \Gamma(\tfrac{1}{4})^2}{r\,\,(9 \pi^{1/2} \lambda^{1/4})} - \frac{2 \sqrt{2} e^{-\tfrac{\gamma}{2} } Q}{\lambda^{1/2}}+\mathcal{O}(r)\,.
\end{eqnarray}
Considering the general case in equations~\eqref{fgencf} and ~\eqref{expfff}, we can compare its expansion with the expansion of the metric function from the logarithmic model in~\eqref{flojEx}. This comparison yields the condition  \( U^{(-1)} = 0 \),  which shows this model does not have a Cauchy horizon.
By comparing the next-order terms, we see that the  expansion self-energy density of the logarithmic theory at \( r = 0 \) is
\begin{eqnarray}\label{selfLoga}
 U^{(0)} =\frac{2 \times 2^{\tfrac{3}{4}} e^{-\tfrac{3}{4}\gamma} Q^{3/2} \Gamma(\tfrac{1}{4})^2}{9 \pi^{1/2} \lambda^{1/4}} ,
\end{eqnarray}
 which corresponds to a finite self-energy. This represents the first demonstration that a theory beyond the Born-Infeld model possesses a finite self-energy for a point charge.
\subsubsection{Thermodynamic of Log-AdS black holes}\label{4222}
\begin{figure}[h]
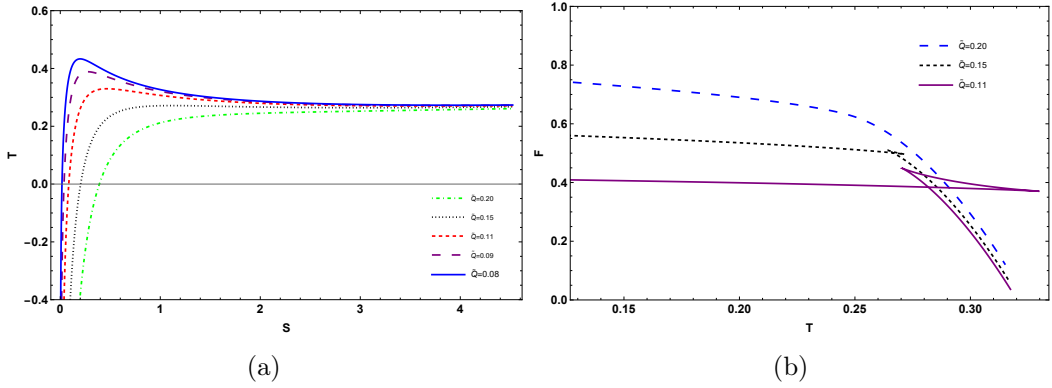

	\begin{subfigure}{0.45\textwidth}\includegraphics[width=\textwidth]{TSLog.pdf}
		\caption{}
		\label{fig:TSlogQ}
	\end{subfigure}
 \begin{subfigure}{0.45\textwidth}\includegraphics[width=\textwidth]{FTLog.pdf}
		\caption{}
		\label{fig:FTlogQ}
	\end{subfigure}
	\caption{ \ref{fig:TSlogQ}: The temperature vs entropy  for Log-$AdS$ black hole for spherical ($k=1 ,L=1$) case by ﬁxed $\lambda=0.02$ for different values of the effective electric charge $\tilde{Q}=e^{-\gamma/2}Q$, and 
    \ref{fig:FTlogQ}: The Helmholtz free energy vs temperature for a Log-$AdS$  black hole for the spherical ($k=1 ,L=1$) case, ﬁxed in  $\lambda=0.02$ for different values of the effective electric charge $\tilde{Q}=e^{-\gamma/2}Q$. }
	\label{fig:FTlog}
\end{figure}

The Hawking temperature, a key thermodynamic quantity, can be derived from the metric solution. Applying Eq.~\eqref{Tempera} to the metric function \(f(r)\) from Eq.~\eqref{fr}, we obtain the temperature as:
\begin{eqnarray}\label{tlogg}
T = \frac{1}{4\pi} \Bigg(\frac{k}{r_h}+\frac{2 r_h}{\lambda}-  r_h \Lambda -  \frac{2 \sqrt{r_h^4 + 2 e^{-\gamma} Q^2 \lambda}}{r_h \lambda}  -  \frac{2 r_h }{\lambda} \log\bigl(\frac{2 r_h^2}{r_h^2 + \sqrt{r_h^4 + 2 e^{-\gamma} Q^2 \lambda}}\bigr)\Bigg).\nonumber\\
\end{eqnarray}
To analyze the thermodynamic stability of the black hole within the logarithmic model, we compute the Helmholtz free energy \( F \), which is the appropriate thermodynamic potential for a system at fixed temperature and volume. We begin with the mass \( M \) of the black hole, defined in~\eqref{Masslogg} as a function of entropy \( S \) and charge \( Q \).  The Helmholtz free energy is defined via a Legendre transformation Eq.~\eqref{F}. Substituting the corresponding expressions for \( M \), \( T \), and \( S \), we obtain:
\begin{eqnarray}\label{freeelog}
F&=& \Omega_{k}\Big(k r_h + \frac{2 r_h^3}{9 \lambda} -  \frac{2 r_h \sqrt{r_h^4 + 2 e^{\gamma} Q^2 \lambda}}{9 \lambda} + \tfrac{1}{3} r_h^3 \Lambda + \frac{32 e^{\gamma} Q^2 }{9 r_h} {}_2F_1(\tfrac{1}{4}, \tfrac{1}{2}, \tfrac{5}{4}, - \tfrac{2 e^{\gamma} Q^2 \lambda}{r_h^4})\nonumber \\
&+& \frac{2 r_h^3 }{3 \lambda} \log\bigl(\frac{2 r_h^2}{r_h^2 + \sqrt{r_h^4 + 2 e^{\gamma} Q^2 \lambda}}\bigr) \Big).
\end{eqnarray}
As shown in Fig.~\eqref{fig:FTlog}, the thermodynamic behavior of gravity coupled to logarithmic nonlinear electrodynamic theory is analogous to that of a van der Waals gas. Similar to the (GBI)theory, the Helmholtz free energy exhibits a characteristic swallowtail structure.

\subsection{Problem to Find an Exact Solution in Causal Theories}\label{05} 
For certain theories, one can determine a compatible potential \( A_\mu \) that permits an analytic solution to Einstein's equations. The aforementioned (GBI) and Logarithmic  (NED) models are specific instances that pertain to this class of solvable systems. Often, even when a suitable potential is identified, solving Einstein's equations is non-trivial, and an analytical solution for the blackening factor remains elusive. This section will address such challenges, with examples drawn from specific theories like the ``No maximum $\tau$'' and ``q-deformed''  theories.
\subsubsection{ No Maximum-$\tau$ Electrodynamics Theory }
\label{51}
In an electrostatic configuration with gauge field \(A_\mu = (A_t(r), 0, 0, 0)\) and no magnetic field, the Lorentz invariant $P$ vanishes, i.e.,\(P = 0\). This simplifies the theory, giving \(U = 0\) and \(V = S\). Consequently, the No Maximum-$\tau$ Lagrangian~\eqref{RootLag} therefore reduces to:
\begin{equation}\label{NoMaxclag}
	\mathcal{L} = -\frac{2}{3\lambda}\left(1 - \left(1 + e^\gamma \lambda S \right)^{3/2}\right).
\end{equation}
Considering the Lagrangian density~\eqref{NoMaxclag},  the antisymmetric tensor \( {G}{_{\mu\nu}} \) is obtained via the derivative of the Lagrangian with respect to the electromagnetic field strength tensor \( {F}{^{\mu\nu}} \):
\begin{equation}\label{RussoGmunuNoMax}
		\mathcal{G}_{\mu\nu} \;=\;e^{\gamma} \sqrt{1 + e^{\gamma} \lambda S}\,  F_{\mu\nu}\,.
\end{equation}
As a result, the energy-momentum tensor \({T}{_{\mu\nu}}\) of the No Maximum \(\tau\) theory will be given by:
\begin{eqnarray}\label{TTnoMax}
T_{\mu \nu} &=& e^{\gamma} F_{\mu }{}^{\alpha } F_{\nu \alpha } \sqrt{1 + e^{\gamma} \lambda S} + \Bigl( \frac{2 (1 + e^{\gamma} \lambda S)^{3/2}}{3 \
\lambda} - \frac{2}{3 \lambda} \Bigr) \mathit{g}_{\mu \nu }.
\end{eqnarray} 
From equation~\eqref{RussoGmunuNoMax}, to find the potential \(A_t(r)\), we need to solve the following differential equation:
\begin{equation}\label{RuNoMax}
	\mathcal{G}_{01} \;=\;D_r=\, - e^{\gamma} \, A^{\prime}_t(r) \sqrt{1 + \frac{1}{2}\,  \lambda\, e^{\gamma} {A^{\prime}_t(r)}^2 }
    \,=\,\frac{Q}{r^2} .
\end{equation}
Analytically solving the governing differential equation~\eqref{RuNoMax} to obtain an explicit form for the potential \(A_t(r)\) leads to an integral with no closed-form solution. Consequently, it is not possible to derive the potential analytically.

This outcome implies that one cannot obtain an exact, analytical solution to the coupled Einstein equations. In the following sections, we introduce a general perturbative approach to circumvent this limitation in the regime of weak nonlinearity. This method enables us to derive black hole solutions for this general theory up to third order in the coupling parameter $\lambda$.
\subsubsection{ $\mathbf{q}$-deform Electrodynamics Theory}
\label{52}
The final closed-form causal nonlinear electromagnetic theory we studied in Section~\eqref{2.5} is the $q$-deformed theory. Here, we discuss why it is impossible for this class of theories to find explicit solutions to Einstein's equations. The Lagrangian can be considered for a purely electric configuration, with no magnetic charge, as follows:
\begin{eqnarray}\label{qDef}
	\mathcal{L}_{\mathbf{q}}=\frac{1}{\lambda} \Big(1 -  (1 - \tfrac{1}{\mathbf{q}} e^{\gamma}\, \lambda \, S )^{\mathbf{q}}\Big)\,. 
\end{eqnarray}
Using Lagrangian~\eqref{qDef}, we can find the antisymmetric  tensors $\mathcal{G}_{\mu\nu}$ as follows:
\begin{equation}\label{RussoGmunuqdef}
		\mathcal{G}_{\mu\nu} \;=\;e^{\gamma} F_{\mu\nu } (1 -  \frac{e^{\gamma} \lambda S}{q})^{q-1}.
\end{equation}
Also, the energy-momentum tensor will be obtained as follows:
\begin{eqnarray}\label{TTnoMax}
T_{\mu \nu} &=& e^{\gamma} F_{\mu }{}^{\alpha } F_{\nu \alpha } {\big(1 -  \frac{e^{\gamma} \lambda S}{q}\big)}^{ q-1} + \frac{1}{\lambda} \Bigl(1 -  {\big(1 -  \frac{1}{q} e^{\gamma} \lambda S\big)}^q\Bigr) \mathit{g}_{\mu \nu}.
\end{eqnarray} 
As is well known, the appropriate potential \( A_t(r) \) must be found by solving the following differential equation:
\begin{equation}\label{Ruqdef}
	\mathcal{G}_{01} \;=\;D_r=\, -  e^{\gamma} A_t^{\prime} (1 -  \frac{ \lambda e^{\gamma} {A_t^{\prime}}^2 }{2 q})^{q-1}
    \,=\,\frac{Q}{r^2} .
\end{equation}
Solving the differential equation~\eqref{Ruqdef} leads to the following solution:
\begin{equation}\label{Aq0}
A_t(r)=\frac{\sqrt{Q \,r_0} \Bigl(( 2 q-1) ( \frac{e^{\gamma} r_0^2 \lambda}{q}-2) + \frac{4 \sqrt{2} (q-1) (2 -  \frac{e^{\gamma} r_0^2 \lambda}{q})^{\tfrac{1+q}{2}}}{2^{\tfrac{q}{2}}}{}_2F_1\bigl(\tfrac{1}{4}, \tfrac{1}{2} (1 + q), \tfrac{5}{4}, \tfrac{e^{\gamma} r_0^2 \lambda}{2 q}\bigr)\Bigr)}{\sqrt{2} e^{ \frac{\gamma}{2} } (2 q-3) (1 -  \frac{e^{\gamma} r_0^2 \lambda}{2 q})^{\tfrac{q}{2}} \sqrt{\frac{e^{\gamma} r_0^2 \lambda}{q}-2}},
\end{equation}
where the value of $r_0$ must be determined from the constraint:
\begin{eqnarray}\label{rq0}
r &=& \frac{\sqrt{Q ( \frac{ \lambda e^{\gamma} r_0^2 }{q}-2)}}{e^{\frac{ \gamma}{2}} \sqrt{2 \, r_0}  (1 -  \frac{e^{\gamma} r_0^2 \lambda}{2 q})^{\tfrac{q}{2}}}.
\end{eqnarray} 
Although the general form of the potential $A_t(r)$ for the $q$-deformed theory is given by Eq.~\eqref{Aq0}, it is in general not possible to determine the value of $r_0$ from Eq.~\eqref{rq0} for arbitrary values of $q$. Only for certain special values, such as $q = 1/2$ and $q = 3/4$, can Eq.~\eqref{rq0} be solved analytically. For instance, the potential for $q = 3/4$ can be obtained, but it remains impossible to solve the corresponding Einstein equations analytically.

These challenges motivate a study of the general behavior of such theories in the presence of gravity via a perturbative expansion of the CH Lagrangian~\eqref{lagg}, using the general Lagrangian given in Eq.~\eqref{Gq1q2n11}. In the following section, we will derive a perturbative solution to the Einstein equations coupled with this (GNED) theory, up to order $\lambda^3$.
We will demonstrate that this perturbative solution captures, to a great extent, the thermodynamic properties and other essential behaviors of black holes arising from the coupling of any electrodynamic theory derived from the Courant-Hilbert approach. Consequently, the analysis of these perturbative solutions is highly valuable.
\subsection{$AdS$ black holes for  (GNED) Theories}\label{06} 
For the (GNED) theory based on coefficients $m_i$ up to $\mathcal{O}(\lambda^3)$ obtained from the Courant-Hilbert expansion in Section~\ref{222.11}, we now analyze the corresponding gravitational solutions.  In this section, we derive solutions to the Einstein equations for this theory by coupling gravity to the general Lagrangian given in~\eqref{Gq1q2n11}. We will show that the expansion of the blackening factor for the GBI and logarithmic models, computed up to order $\lambda^3$, is consistent with the blackening factor obtained from the (GNED) framework.

Considering only the electric charge $Q$ and assuming the absence of magnetic charge, the $\lambda^4$-order expansion of the CH function in Eq.~\eqref{Lexpan} yields the following Lagrangian term:
\begin{eqnarray}\label{lags}
\mathcal{L}_{GNED} &=&e^{\gamma} S + m_1 \lambda e^{2 \gamma} S^2 +m_2 \lambda^2 e^{3 \gamma}  S^3 + m_3 \lambda^3 e^{4 \gamma} S^4 +m_4 \lambda^4 e^{5 \gamma}  S^5+\cdots .
\end{eqnarray} 
The antisymmetric tensor $\mathcal{G}_{\mu \nu}$ is derived from the (GNED) Lagrangian in \eqref{lags},  by taking its derivative with respect to the electromagnetic field strength tensor. This leads to a specific expression for the  $\mathcal{G}_{01}$ tensor in (GNED) theory by $A_{\mu}=(A_t(r),0,0,0)$. Finally, to determine the electric potential $A_t(r)$, one must solve a differential equation that comes from combining two of the previously established relations. This differential equation is as follows:
\begin{equation}\label{RuGNED}
	\mathcal{G}_{01} =- e^{\gamma} {A_t^{\prime}}-  m_1  \lambda \,e^{2 \gamma}  {A_t^{\prime}}^3  - \tfrac{3}{4} m_2 \lambda^2 e^{3 \gamma} {A_t^{\prime}}^5  - \tfrac{1}{2}   m_3 \lambda^3   e^{4 \gamma} {A_t^{\prime}}^7-  \tfrac{5}{16} m_4  \lambda^4  e^{5 \gamma}   {A_t^{\prime}}^9
    \,=\,\frac{Q}{r^2} .
\end{equation}
By perturbation approach, solving the governing differential equation~\eqref{RuGNED} to find an explicit form for the potential $A_t(r)$ leads to a perturbation solution as:
\begin{eqnarray}\label{AtGen}
A_t(r)  &=&e^{-\gamma} \frac{ Q}{r}  
		-  \lambda\, e^{-2 \gamma} \frac{ m_1 Q^3}{5\, r^5} 
		+ \lambda^2  e^{-3 \gamma}\frac{ (4 m_1^2 -  m_2) Q^5 }{12\, r^9} - \lambda^3 e^{-4 \gamma}  \frac{(24 m_1^3 - 12 m_1 m_2 + m_3) Q^7}{26 \,r^{13}} \nonumber\\
&+& 5\lambda^4 e^{-5 \gamma}  \frac{5 (176 m_1^4 - 132 m_1^2 m_2 + 9 m_2^2 + 16 m_1 m_3 -  m_4) Q^9 }{272\,\, r^{17}}\, .
\end{eqnarray} 
The energy-momentum tensor for (GNED) theory  will be obtained as follows:
\begin{eqnarray}\label{TTGNED}
T_{\mu \nu} &=&(e^{\gamma} + 2 m_1 \lambda e^{2 \gamma}   S + 3 m_2 \lambda^2e^{3 \gamma}   S^2 + 4m_3 \lambda^3  e^{4 \gamma} S^3 + 5  m_4 \lambda^4 e^{5 \gamma}  S^4)F_{\mu }{}^{\alpha } F_{\nu\alpha } \\
&+& (e^{\gamma} S + m_1 \lambda e^{2 \gamma} S^2 +m_2 \lambda^2 e^{3 \gamma}  S^3 + m_3 \lambda^3 e^{4 \gamma} S^4 +m_4 \lambda^4 e^{5 \gamma}  S^5) \mathit{g}_{\mu \nu}.\nonumber
\end{eqnarray} 
Solving Einstein’s equations~\eqref{EinE} together with the energy–momentum tensor~\eqref{TTGNED} leads to the following differential equations:
\begin{eqnarray}\label{T00T11}
&&f(r)(- \Lambda -  \frac{f^{\prime}(r)}{r})  -  \frac{{f(r)}^2}{r^2}-  \frac{e^{-\gamma}  Q^2}{r^4}f(r)+\lambda \frac{e^{-2 \gamma} m_1 Q^4 }{2 r^8} f(r) \\
&& -\lambda^2  \frac{e^{-3 \gamma} (4 m_1^2 -  m_2) Q^6 }{4 r^{12}}f(r)+  \lambda^3\frac{e^{-4 \gamma} (24 m_1^3 - 12 m_1 m_2 + m_3) Q^8}{8 r^{16}} f(r)\nonumber\\
&&+\lambda^4\frac{e^{-5 \gamma}  (176 m_1^4 - 132 m_1^2 m_2 + 9 m_2^2 + 16 m_1 m_3 -  m_4) Q^{10} }{16\, r^{20}}f(r)=0,\nonumber
\end{eqnarray}
and
\begin{eqnarray}\label{T022}
&& \frac{1}{2} (f^{\prime \prime}(r) + 2 \Lambda) r^2  + f^{\prime}(r)\, r  -  \frac{e^{-\gamma} Q^2}{r^2}+\lambda \frac{3 e^{-2 \gamma} m_1 Q^4 }{2\, r^6}\\
&& - \lambda^2 \frac{5 e^{-3 \gamma} (4 m_1^2 -  m_2) Q^6 }{4 \,r^{10}}+\lambda^3 \frac{7 e^{-4 \gamma} (24 m_1^3 - 12 m_1 m_2 + m3) Q^8 }{8 \,r^{14}}\nonumber \\
&&+\lambda^4\frac{9 e^{-5 \gamma} (176 m_1^4 - 132 m_1^2 m_2 + 9 m_2^2 + 16 m_1 m_3 -  m_4) Q^{10} }{16 r^{18}} =0.\nonumber
\end{eqnarray}
Solving the above Einstein’s field equations, together with the metric ansatz~\eqref{HigherDMetric}, one obtains the following blackening factor:
\begin{equation}\label{BFactorgned}
	\begin{aligned}
		f(r) = &\, k -  \frac{1}{3} r^2 \Lambda -  \frac{m}{r}
		+ e^{-\gamma}  \frac{ Q^2}{r^2} 
		- \lambda\, e^{-2 \gamma}   \frac{m_1 Q^4 }{10 r^6}  
		\\
		&+ \lambda^2\,  e^{-3 \gamma}\frac{ (4 m_1^2 -  m_2) Q^6}{36 r^{10}} - \lambda^3 \,e^{-4 \gamma} \frac{ (24 m_1^3 - 12 m_1 m_2 + m_3) Q^8 }{104 r^{14}} \\
		& +\lambda^4\frac{ e^{-5 \gamma} (176 m_1^4 - 132 m_1^2 m_2 + 9 m_2^2 + 16 m_1 m_3 -  m_4) Q^{10} }{272\, r^{18}}  \, .
	\end{aligned}
\end{equation}
An analysis of the black hole horizon within the perturbation approach depends critically on the order of the perturbation and the specific theory, which is determined by the $m_i$ coefficient dependencies. Depending on these factors, the black hole metric can exhibit single-horizon, two-horizon, or multi-horizon behavior \eqref{BFactorgned}.
The horizon radius $r_h$, defined as the largest root of $f(r_h)=0$, is related to the mass parameter $m$ through
\begin{equation}\label{horizgned}
	\begin{aligned}
		m = &\, k\, r_h - \frac{1}{3} r_h^3 \Lambda  
		+ e^{-\gamma}\,\frac{ Q^2}{r_h} 
		- \lambda \,  e^{-2 \gamma}\,  \frac{m_1 \, Q^4 }{10 r_h^5}  \\
		& + \lambda^2 \, e^{-3 \gamma}\frac{ (4 m_1^2 -  m_2) Q^6}{36 r_h^9}
		- \lambda^3\,e^{-4 \gamma} \,\frac{(24 m_1^3 - 12 m_1 m_2 + m_3)\, Q^8 }{104 r_h^{13}} \\
		& +\lambda^4 e^{-5 \gamma}  \frac{ (176 m_1^4 - 132 m_1^2 m_2 + 9 m_2^2 + 16 m_1 m_3 -  m_4) Q^{10} }{272\, r_h^{17}}  \, .
	\end{aligned}
\end{equation}
Consequently, the mass of the black hole is $M = 2\Omega_{k}m$.

A central result of this paper is the blackening factor~\eqref{BFactorgned}, which provides a unified framework for studying electrodynamic theories in curved spacetime. By substituting the expansion coefficients $m_i$ into this metric, as summarized in Table~\eqref{tab1}, one can systematically analyze the corresponding thermodynamic properties. This method successfully reproduces the behavior of GBI and logarithmic electrodynamics, and further enables the study of Maxwellian deformations, $q$-deformed models, and cases with no maximum $\tau$, for which closed-form solutions to Einstein’s equations were previously inaccessible. The blackening factor~\eqref{BFactorgned} therefore establishes a consistent setting for investigating the thermodynamics of all deformed Maxwell theories.

\subsubsection{Thermodynamic of GNED-AdS black holes}\label{509}
The Hawking temperature at the event horizon can be obtained from Eqs.~\eqref{Tempera} and~\eqref{BFactorgned}, which yield
\begin{eqnarray}\label{Ttned}
T &=&\frac{1}{4\pi} \Bigg( \frac{k}{r_h}-  r_h \Lambda   
		- e^{-\gamma} \frac{ Q^2}{r_h^3} 
		+\lambda e^{-2 \gamma} \frac{ m_1 Q^4 }{2 r_h^7} 
		\\
        &+& \lambda^2 e^{-3 \gamma} \frac{ (-4 m_1^2 + m_2) Q^6 }{4 r_h^{11}}+\lambda^3 e^{-4 \gamma}\frac{ (24 m_1^3 - 12 m_1 m_2 + m_3) Q^8 }{8 r_h^{15}} \nonumber\\
		&-&\lambda^4 e^{-5 \gamma}  \frac{ (176 m_1^4 - 132 m_1^2 m_2 + 9 m_2^2 + 16 m_1 m_3 -  m_4) Q^{10} }{16\, r_h^{19}} \Bigg). \nonumber
\end{eqnarray}

\begin{figure}
	\centerline{\includegraphics[scale=.6]{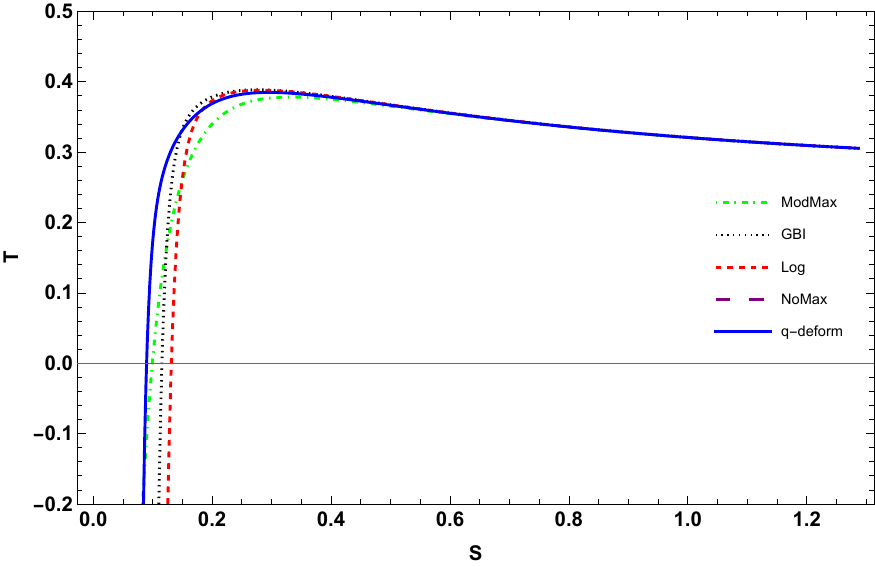}}
	\caption{The inverse Hawking  temperature vs entropy for different theories for spherical ($k=1 ,L=1$) case by ﬁxed $\lambda=0.02$ and  $e^{-\gamma/2} \,Q=0.09$. Here we considered $q=3/5$ for the $q$-deformed theory. }\label{Tall}
\end{figure}

\begin{figure}
	\centerline{\includegraphics[scale=.6]{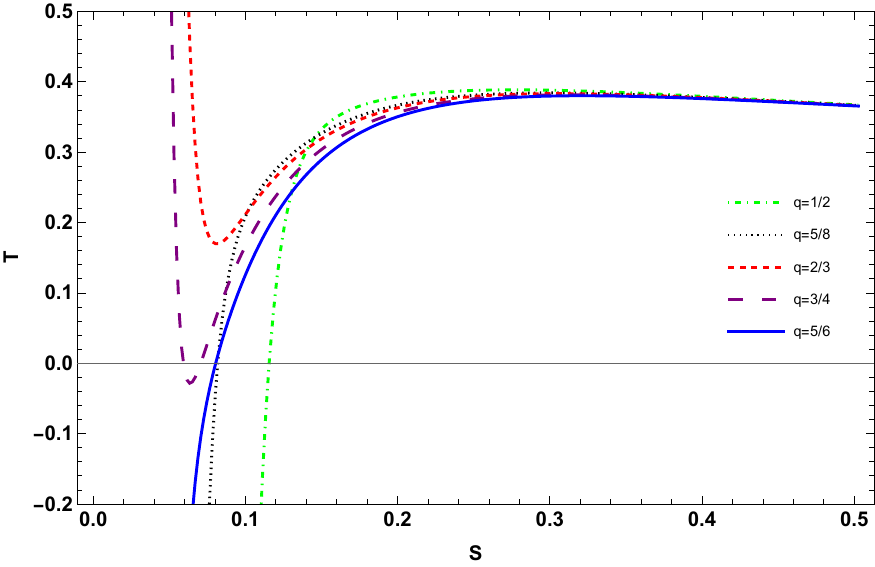}}
	\caption{The inverse Hawking  temperature vs entropy for for the $q$-deformed  theories for spherical ($k=1 ,L=1$) case by ﬁxed $\lambda=0.02$ and  $e^{-\gamma/2} \,Q=0.09$. }\label{q-defooo}
\end{figure}

By considering perturbations  up to order $\lambda^4$, we observe thermodynamic behavior that is universal across all the theories studied. This shared behavior closely mirrors that of the GBI and logarithmic models in their non-perturbative regimes. However, since all quantities depend on the parameters \( m_i \), their specific behavior may differ across theories for different choices of the set \( \{m_i\} \). To substantiate this observation, we compare the temperature profiles of these theories. Fig.~\eqref{Tall} displays temperature as a function of entropy, showing that all theories exhibit nearly identical behavior at the order $\lambda^4$ . This behavior is fully consistent with the temperature profiles of the (GBI) and logarithmic black holes shown in Figs.~\eqref{fig:TSBIQ} and~\eqref{fig:TSlogQ}.
It is noteworthy that in the ModMax theory, the term $e^{-\gamma/2} Q$ may be treated as an effective charge, since the constant factor $e^{-\gamma/2}$ can be absorbed into $Q$. With this rescaling, the ModMax T-S curve retains the same qualitative behavior and reproduces the Maxwell-like behavior.
 It would be instructive to perform a temperature analysis of the $q$-deformed theories. 
In Figure~\ref{q-defooo}, we plot temperature versus entropy for several values of $q$. To ensure the causality condition is satisfied, we have considered the range $1/2 \leq q \leq 1$. The resulting curves reveal two distinct qualitative behaviors. For $q = 2/3$ and $q = 3/4$, the temperature exhibits a noticeably different profile near the low-entropy peak compared to other values of $q$. However, at large entropies, all curves converge and saturate to the same asymptotic temperature.

\begin{figure}[h]
	\begin{subfigure}{0.45\textwidth}\includegraphics[width=\textwidth]{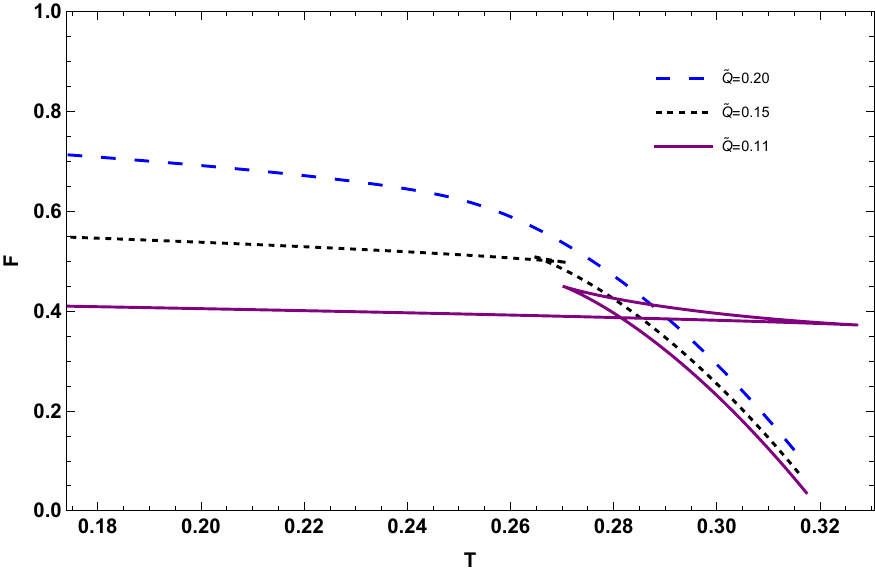}
		\caption{}
		\label{nomafreeEE}
	\end{subfigure}
 \begin{subfigure}{0.45\textwidth}\includegraphics[width=\textwidth]{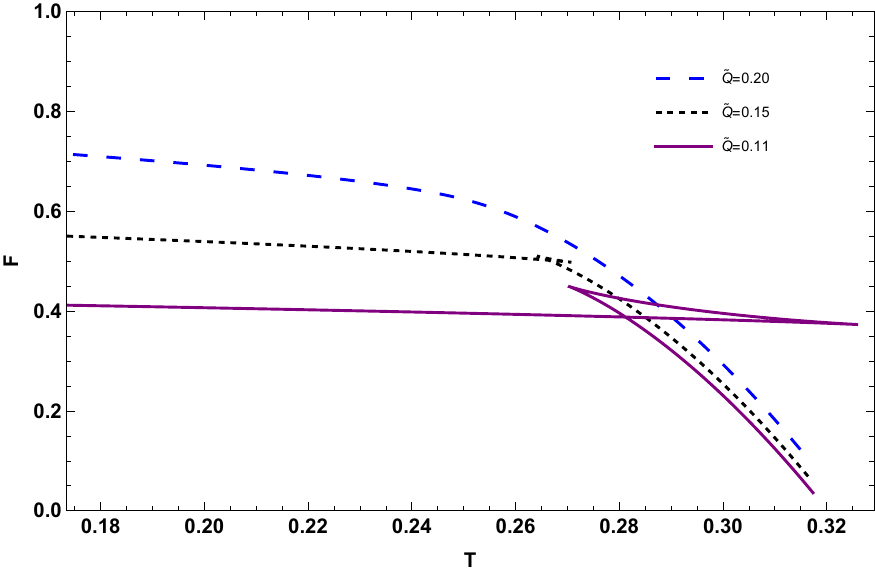}
		\caption{}
		\label{qdefferrr}
	\end{subfigure}
	\caption{ \ref{nomafreeEE}: The Helmholtz free energy vs temperature for a No Max-$\tau$-$AdS$  black hole for the spherical ($k=1 ,L=1$) case, ﬁxed in  $\lambda=0.01$ for different variable effective electric charge $\tilde{Q}=e^{-\gamma/2}Q$, and 
    \ref{qdefferrr}: The Helmholtz free energy vs temperature for a $q$-deformed-$AdS$  black hole for the spherical ($k=1 ,L=1$) case, ﬁxed in  $\lambda=0.01$ and $q=3/4$ for different variable effective electric charge $\tilde{Q}=e^{-\gamma/2}Q$. }
	\label{NOMAxqdefFreeE}
\end{figure}

To analyze the thermodynamic stability of a black hole in (GNED), we calculate the Helmholtz free energy $F$, which represents the appropriate thermodynamic potential for  systems at fixed temperature and volume. By examining the free energy, we can investigate possible phase transitions between small and large black holes across different models. Substituting the expressions for mass $M$, temperature $T$, and entropy $S$  of GNED-AdS black holes, into Eq.~\eqref{F}, we have:
\begin{eqnarray}\label{Frenedg}
F &=& k \,r_h + \frac{1}{3} r_h^3 \Lambda + e^{-\gamma}\frac{3 Q^2}{r_h} 
		- \lambda e^{-2 \gamma} \frac{7 m_1 Q^4 }{10 r_h^5} 
		\\
        &+& + \lambda^2 e^{-3 \gamma}\frac{11 (4 m_1^2 - m_2) Q^6 }{36 r_h^9} 
		- \lambda^3 e^{-4 \gamma} \frac{15(24 m_1^3 - 12 m_1 m_2 + m_3) Q^8 }{104 r_h^{13}} \nonumber\\
		&+&\lambda^4 e^{-5 \gamma}  \frac{19 (176 m_1^4 - 132 m_1^2 m_2 + 9 m_2^2 + 16 m_1 m_3 -  m_4) Q^{10} }{272\, r_h^{17}}.  \nonumber
\end{eqnarray}
With the Helmholtz free energy given in Eq.~\eqref{Frenedg}, we are now able to study phase transitions across the different models. The behavior of the above free energy  for both the Born-Infeld and logarithmic theories reproduces exactly the results obtained in Sections~\eqref{4111} and~\eqref{4222}. Therefore, we do not repeat hose results here. Instead, we focus on analyzing the free energy in Eq.~\eqref{Frenedg} as a function of temperature for the  No Maximum-$\tau$ theory and the $q$-deformed theories. As expected, the characteristic swallowtail behavior appears in both the no-maximum model and the $q$-deformed  model, as shown in Figures~\eqref{nomafreeEE} and~\eqref{qdefferrr}. Our analysis demonstrates that with fixed values of $e^{-\gamma/2}\,Q$ and $\lambda$ for $(k=1, L=1)$, the swallowtail structure remains qualitatively unchanged when varying the parameter interval. Consequently, we show only a single representative case of the $q$-deformed models for $q = 3/4$.

\section{Kretschmann scalar and Singularities of black holes}\label{055}
In this section, we propose that the structure of this singularity is not arbitrary but is intrinsically determined by the fundamental properties of the matter that formed the black hole. 
The Kretschmann scalar is defined as the full contraction of the Riemann tensor:
\[
\mathcal{K} \equiv R_{\mu\nu\rho\sigma}R^{\mu\nu\rho\sigma}.
\]
The Kretschmann scalar $\mathcal{K}$ for a charged, non-rotating black hole demonstrates a true curvature singularity at the origin $r=0$. The strength of this singularity is intimately connected to the electromagnetic source of the geometry.
For a static, spherically symmetric metric of the general form~\eqref{HigherDMetric}, the Kretschmann scalar constructed from the Riemann tensor reads
\begin{equation}\label{KH}
\mathcal{K}= f''(r)^{2} + \frac{4 f'(r)^{2}}{r^{2}} + \frac{4 f(r)^{2}}{r^{4}} - \frac{8 k f(r)}{r^{4}} + \frac{4 k^{2}}{r^{4}},
\end{equation}
where the behavior of \(f(r)\) near the origin is governed by the local energy density of the electromagnetic field.
\subsection{Singularity of ModMax black hole}
In the limit  $\lambda = 0$, the Genrealized Nonlinear Electrodynamic Anti-de Sitter (GNED-AdS) black hole solution reduces to that of the ModMax-AdS black hole~\cite{Flores-Alfonso:2020euz,BallonBordo:2020jtw,Bokulic:2021dtz,Pantig:2022gih,Barrientos:2022bzm,Kruglov:2021bhs,Kruglov:2022qag,Barrientos:2024umq,Guzman-Herrera:2023zsv,Zhang:2021qga,EslamPanah:2024gxx,EslamPanah:2024qlu,Heidari:2025llu}. The metric for this limiting case for $k=1$ is given by:
\[
ds^2 = -\left(1 - \frac{m}{r} +e^{-\gamma} \frac{Q^2}{r^2}\right) dt^2 + \left(1 - \frac{m}{r} +e^{-\gamma}  \frac{Q^2}{r^2}\right)^{-1} dr^2 + r^2 d\Omega_k^2.
\]
For the ModMax black hole, an explicit calculation yields:
\begin{equation}
\mathcal{K} = \frac{56\, e^{-2\gamma} Q^4}{r^8} - \frac{48 \,m\, e^{-\gamma} Q^2}{r^7} + \frac{12\, m^2}{r^6}.
\label{eq:kretschmann}
\end{equation}

\subsection*{ Analysis of the Singularity at $r \to 0$}

As we approach the singularity, the dominant term in Eq.~\eqref{eq:kretschmann} is the one with the highest power of $1/r$:
\[
\mathcal{K} \sim \frac{56 e^{-2\gamma}Q^4}{r^8} \quad \text{as} \quad r \to 0.
\]
This indicates a mathematical singularity of order $1/r^8$. The appearance of a "$1/r^6$ singularity"  reflects a more nuanced physical interpretation.

\begin{itemize}
    \item \textbf{The $1/r^8$ Term:} This term, $\propto Q^4/r^8$, originates directly from the electromagnetic field. A point charge has an electric field $E \propto 1/r^2$. The electromagnetic stress-energy tensor $T_{\mu\nu}^{\text{(EM)}}$ is quadratic in the field strength, $F_{\mu\nu}F^{\mu\nu} \propto E^2 \propto 1/r^4$. Since Einstein's equations, $G_{\mu\nu} = 8\pi T_{\mu\nu}$, link curvature to the stress-energy tensor, it is natural for a curvature invariant like $\mathcal{K}$ to scale as $(1/r^4)^2 = 1/r^8$.

    \item \textbf{The $1/r^6$ Term:} This term, $12 m^2 / r^6$, is the \textit{only} singularity present in the uncharged Schwarzschild black hole ($Q=0$). It represents the pure gravitational curvature singularity associated with a point mass.
\end{itemize}
\subsection{ Naked singularity of $AdS$-GBI black hole}
 Studies in~\cite{Chen:2023trn,Guo:2022ghl} investigate black hole solutions and naked Born-Infeld singularities within a framework of gravity coupled to Born-Infeld electrodynamics. These works show that when the singularity exhibits a specific, well-ordered structure, it allows photons entering the photon sphere(s) to pass through it.
We now proceed to examine the singularity structure of the  (GBI) theory. A distinctive feature of this model, as derived from its Lagrangian, is the emergence of an energy–density singularity that scales as as $1/r^6$. For the electrically charged black holes considered, the parameter $\gamma$ only appears in the combination $e^{-\gamma/2}Q$. This implies that the physical behavior in the (GBI) and (GNED) theories is identical to the case with $\gamma = 0$, as it is governed entirely by a single effective charge $\tilde{Q} = e^{-\gamma/2}Q$. This behavior is characterized by the parameter~$\gamma$, which is intrinsic to the non-linear electrodynamics of the GBI action. To understand the gravitational implications, we analyze the metric function~$f(r)$ obtained by solving the coupled Einstein-GBI field equations. Our objective is to determine how this $\mathcal{O}(1/r^6)$ singularity, governed by~$\gamma$, manifests itself in the geometry near the origin~$r \to 0$ and influences the global causal structure of the spacetime.
Using the definition of the Kretschmann scalar in \eqref{KH} and the  function \(f(r)\) in \eqref{frGBI}, we compute the series expansion of Kretschmann  scalar at \(r= 0\) for AdS-GBI   as follows:
\begin{equation}\label{KGBI}
	\mathcal{K} = 12 \biggl(m -  \frac{ e^{-3/4 \gamma} Q^{3/2} \Gamma(\tfrac{1}{4})^2}{3 \pi^{1/2} \lambda^{1/4}}\biggr)^2 r^{-6} +\mathcal{O}(r^{-5}).
\end{equation}
A spacetime configuration represents a black hole solution if and only if it possesses an event horizon. The critical boundary that distinguishes a black hole from a naked singularity is defined by the existence of an extremal black hole. To identify this separatrix in parameter space, we analyze the extremal case at $\Lambda=0$, which has a degenerate horizon at radius \( r_e \) and a corresponding mass \( M_e =2 \, m_e \,\Omega_k\). The defining conditions for this extremal configuration are:
\begin{equation}\label{exter}
\left. \frac{d}{dr} \,\big(r\, f(r)\big) \right|_{r = r_e}=0\,,\qquad f(r_e)=0.
\end{equation}
From Condition~\eqref{exter}, we can find the extremal radius \( r_e \) and the parameter \( m_e \) as follows:
\begin{equation}\label{reee}
r_e=\frac{ \sqrt{k} }{2}\sqrt{\frac{4 e^{-\gamma} Q^2}{k^2} -  \lambda}; \quad 
m_e=\frac{1}{3} k^{3/2} \sqrt{\frac{4 e^{\gamma} Q^2}{k^2} -  \lambda}\,+\frac{8 e^{-\gamma} Q^2 \,{}_2F_1\bigl(\tfrac{1}{4}, \tfrac{1}{2}, \tfrac{5}{4}, - \frac{16 e^{-\gamma} Q^2 \lambda}{k^2 (\lambda- \frac{4 e^{-\gamma} Q^2}{k^2} )^2}\bigr)}{3 \sqrt{k} \sqrt{\frac{4 e^{-\gamma} Q^2}{k^2} -  \lambda}}
\end{equation}
It is evident that extremal black holes do not exist for $k = -1$ or $k = 0$. Therefore, we restrict our analysis to the case \( k= 1 \). Substituting this value into the field equations, the condition for the existence of a real and positive extremal radius \( r_e \) yields a direct constraint among the parameters, namely \( \lambda < 4 e^{-\gamma} Q^2\). Consequently, for masses below the extremal bound, $M < M_e$, the spacetime represents a naked singularity.
Furthermore, the conditions \( \frac{d}{dr} \,\big(r\, f(r)\big) > 0 \) and
\begin{equation}\label{limoo}
\displaystyle \lim_{r \to 0} \,r\, f(r)=- m + \frac{ e^{-\tfrac{3}{4}\gamma} Q^{3/2} \Gamma(\tfrac{1}{4})^2}{3 \pi^{1/2} \lambda^{1/4}}
\end{equation}
together imply the presence of a naked singularity, which occurs when:
\begin{equation}\label{limoo}
m \,<\, \frac{ e^{-\tfrac{3}{4}\gamma} Q^{3/2} \Gamma(\tfrac{1}{4})^2 }{3 \pi^{1/2} \lambda^{1/4}}.
\end{equation}
From the self-energy in Eq.~\eqref{SelfEBI}, we see that the condition $m < U^{(0)}$ leads to a naked singularity in the black hole.
\subsection{Naked singularity of $AdS$-Log black hole}
The series expansion of the Kretschmann scalar for the AdS-Log theory at \(r=0\), derived from \eqref{KH} and the metric  function  \eqref{fr}, is given by:
\begin{eqnarray}\label{KLogsin}
	\mathcal{K} = 12 \biggl(m -  \frac{2 \times 2^{\tfrac{3}{4}} e^{-\tfrac{3}{4} \gamma} Q^{3/2} \Gamma(\tfrac{1}{4})^2}{9 \pi^{1/2} \lambda^{1/4}}\biggr)^2 \, r^{-6} +\mathcal{O}(r^{-5}).
\end{eqnarray}
 We expand \( r f(r) \) around \( r=0 \) in order to determine the parameter intervals that lead to a naked singularity. The expansion takes the form:

\begin{equation}\label{limlogo}
r\, f(r)=- m +  \frac{2 \times 2^{\tfrac{3}{4}} e^{-\tfrac{3}{4} \gamma} Q^{3/2} \Gamma(\tfrac{1}{4})^2}{9 \pi^{1/2} \lambda^{1/4}}+\Big( k-\frac{2\sqrt{2} e^{-\tfrac{\gamma}{2}}Q} {\sqrt{\lambda}}\Big) \,r +\cdots .
\end{equation}
While the extremality condition in \eqref{exter} cannot be solved analytically for logarithmic black holes, the nature of the singularity can be investigated via the condition \( \frac{d}{dr} \,\big(r\, f(r)\big) > 0 \).
\begin{equation}\label{limoggllo}
\frac{d}{dr} \,\big(r\, f(r)\big) = k-\frac{2\sqrt{2} e^{-\tfrac{\gamma}{2}}Q} {\sqrt{\lambda}} +\cdots .
\end{equation}
As evident from \eqref{limoggllo}, the condition $\frac{d}{dr} \,\big(r\, f(r)\big) > 0$ holds only in the case of \( k = 1 \), under the following parameter restriction:
\begin{equation}
\lambda >  8 e^{-\gamma} Q^2
\end{equation}
For \(\lambda >  8 e^{-\gamma} Q^2 \), the spacetime admits at most one horizon. The existence of this horizon is determined by examining the function \( r f(r) \), which vanishes at the horizon radius.
Furthermore, taking   \( \displaystyle \lim_{r \to 0} \,r\, f(r)\), we find:
\begin{equation}\label{limggho}
\displaystyle \lim_{r \to 0} \,r\, f(r)=- m + \frac{2 \times 2^{\tfrac{3}{4}} e^{-\tfrac{3}{4} \gamma} Q^{3/2} \Gamma(\tfrac{1}{4})^2}{9 \pi^{1/2} \lambda^{1/4}}.
\end{equation}
The simultaneous satisfaction of these criteria points to the existence of a naked singularity in Log-$AdS$ black holes, a scenario that arises when:
\begin{equation}\label{limoergolo}
m\,<\, \frac{2 \times 2^{\tfrac{3}{4}} e^{-\tfrac{3}{4} \gamma} Q^{3/2} \Gamma(\tfrac{1}{4})^2}{9 \pi^{1/2} \lambda^{1/4}}.
\end{equation}
For the logarithmic theory, we find from the self-energy in Eq.~\eqref{selfLoga} that the condition $m < U^{(0)}$ leads to the formation of a naked singularity.

\begin{figure}[h]
	\begin{subfigure}{0.45\textwidth}\includegraphics[width=\textwidth]{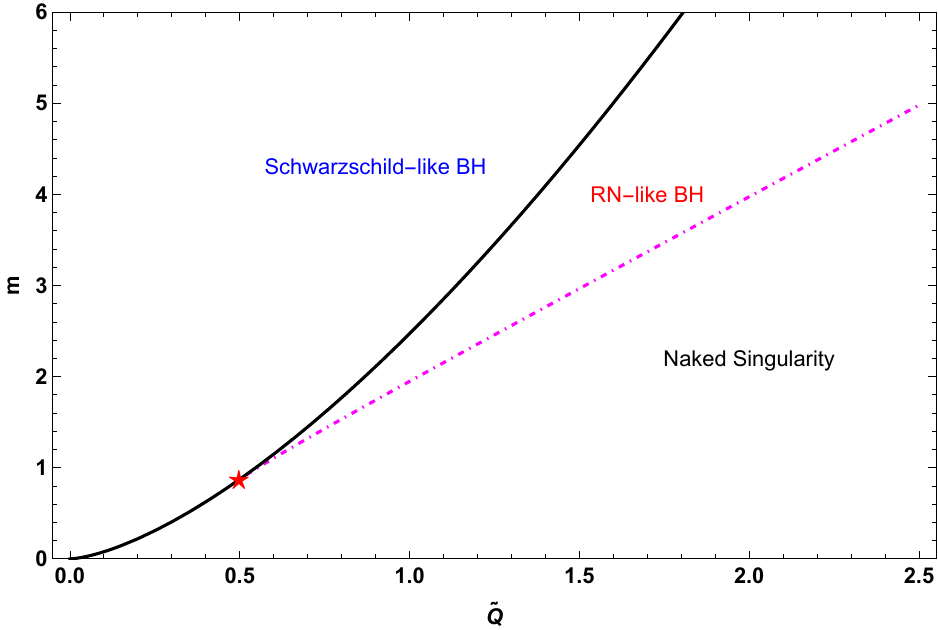}
		\caption{}
		\label{NSingGBI}
	\end{subfigure}
 \begin{subfigure}{0.45\textwidth}\includegraphics[width=\textwidth]{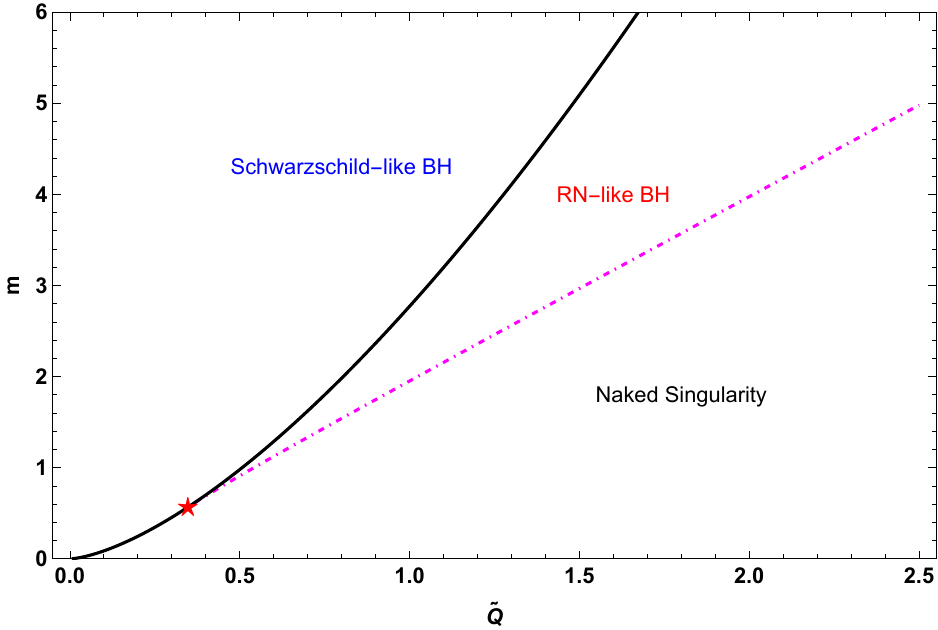}
		\caption{}
		\label{NSinLog}
	\end{subfigure}
	\caption{ Parameter space $(\tilde{Q}, m)$ for static Born-Infeld black holes in Fig.~\ref{NSingGBI} and for the logarithmic electrodynamics theory in Fig.~\ref{NSinLog}.  The $\tilde{Q}=e^{-\gamma/2}Q$ is a effective charge.  The thick black curve corresponds to Schwarzschild-like black holes, while the region between this curve and the magenta dot-dashed curve represents Reissner-Nordström-like black holes. The magenta dot-dashed curve describes extremal black holes; it terminates at the intersection with the thick black curve-the `star point'. Here we have set $k=1$, $\Lambda=0$, and fixed the coupling $\lambda=1$.}
	\label{NOMAxqdefFreeE}
\end{figure}

The panel in Fig.~\eqref{NSingGBI} illustrates the phase space of solutions as a function of the charge-to-mass ratio and the mass, with the parameter fixed according to Eqs.~\eqref{reee} and~\eqref{limoo} for (GBI) theory. Similar behavior is shown for the logarithmic theory, with its extremal mass plotted via a numerical method in Fig.~\eqref{NSinLog}. The star point in (GBI) black hole corresponds exactly to the limit \( \lambda = 4 e^{-\gamma} Q^2 \), while in logarithmic black hole, the star point corresponds to the limit \( \lambda = 8 e^{-\gamma} Q^2\) for \( \lambda=1 \).

We have shown in this section that the singularity structure of black holes is deeply tied to the properties of the matter fields that source the geometry, and the Kretschmann scalar provides a precise measure of the curvature strength at the center. For charged black holes, this quantity represents the true singularity at the origin, whose behavior depends on the electromagnetic contribution: in the ModMax theory, the curvature grows strongly due to the charge term, although the dominant contribution arises from gravity, reproducing the strong singularity of the usual Schwarzschild type. In nonlinear electrodynamics models such as the  (GBI) and logarithmic theories, the singularity inherits characteristic features of the underlying Lagrangian, producing dominant divergences that scale with fixed powers of the radial coordinate determined by the parameters of the theory. By analyzing the near-origin expansions and the behavior of the metric function, one can identify the conditions under which an event horizon forms; failure to satisfy these conditions leads to naked singularities whose existence is controlled by relations among the mass, charge, curvature parameter, and Lagrangian couplings. This comparative analysis demonstrates how different causal nonlinear electrodynamics theories imprint distinct singularity structures and horizon formation criteria on their corresponding black hole solutions.

 \section{Discussion and Conclusions}\label{Discuss}

In this work, we have developed a unified and causal framework for analyzing charged black holes in nonlinear electrodynamics (NED) coupled to Einstein gravity, where the electromagnetic sector is governed by a root-$T\bar{T}$ deformation of duality-invariant field theories.
The resulting (GNED) model encapsulates a broad class of theories, including ModMax, (GBI),  self-dual logarithmic electrodynamics,  No Maximum-$\tau$, and $q$-deformed theories as specific limits of its coupling parameters. 

The thermodynamic analysis of the resulting black hole solutions reveals a rich phase structure. The free-energy landscape exhibits the characteristic swallowtail structure associated with first-order phase transitions between small and large black holes, reminiscent of the van der Waals behavior in standard thermodynamics. The influence of the $T\bar{T}$ and root-$T\bar{T}$ deformations appears through the modified free energy and specific heat, both of which are directly controlled by the (NED)  couplings. These deformations shift the location of the critical point and modify the stability domains of the black hole phases, leading to a tunable thermodynamic response that smoothly interpolates between linear (Maxwell) and nonlinear regimes.

The central singularity of a black hole is closely tied to the matter fields that generate its spacetime, with the Kretschmann scalar providing a precise measure of how strongly the curvature diverges at the origin. For charged black holes, including those in ModMax theory, the singularity reflects both electromagnetic and gravitational contributions, though the dominant divergence is set by the purely gravitational term inherited from the Schwarzschild solution. In nonlinear electrodynamics (NED) models such as the  (GBI) and logarithmic theories, the nonlinearity of the electromagnetic Lagrangian modifies the high-energy behavior of the field, leaving a direct imprint on the geometry and determining how sharply the curvature grows near the center. These models often produce characteristic singularity profiles that differ from the Maxwell case, reflecting the theory’s ultraviolet structure. By examining the metric near the origin, one can identify precise conditions under which an event horizon forms and hides the singularity; violation of these conditions leads to naked singularities, whose existence is governed by specific relations among the mass, charge, and coupling constants. Thus, different nonlinear electrodynamics theories yield distinct singularity structures and horizon formation criteria, highlighting the deep interplay between matter content and spacetime geometry.

The work by Russo and Townsend~\cite{Russo:2025fuc} provides a novel formulation of self-dual nonlinear electrodynamics where interactions are determined by an auxiliary-field potential. A central achievement is the resolution of a long-standing problem: the derivation of an explicit Lagrangian for the generic `analytic' theory, which is accomplished by restricting to even potential functions. The Russo and Townsend framework proves powerful for analyzing black hole solutions, yielding new perspectives on their singularities and overall spacetime structure~\cite{Russo:2026vnj}.

Holographically~\cite{Witten:1998qj}, one could use the AdS/CFT correspondence~\cite{Maldacena:1997re} to compute how GNED black holes modify transport coefficients and probe quantum information observables like entanglement entropy~\cite{Ryu:2006bv,Doi:2022iyj,MohammadiMozaffar:2016vpf,BabaeiVelni:2019pkw} and complexity~\cite{Alishahiha:2015rta,Brown:2015lvg,Yekta:2020wup,Babaei-Aghbolagh:2021ast}, revealing nonlinear effects on strongly coupled systems. 
Topologically thermodynamic defects~\cite{Wei:2022dzw}, applying modern classification methods to the black hole's phase structure could uncover new stability patterns and universal classes dictated by the nonlinear electrodynamics~\cite{Yerra:2022eov,NooriGashti:2025hpz,Babaei-Aghbolagh:2025qxm}. 
Collectively, these holographic, information-theoretic, and topological investigations would provide a deep and unified understanding of the GNED model's role within causal, duality-consistent theories of gravity and nonlinear electrodynamics.

Finally, our framework provides a promising foundation for future work on holographic interpretations of $T\bar{T}$ and root-$T\bar{T}$ flows in AdS gravity. The causal and duality-preserving character of the (GNED) sector suggests that these deformations may encode novel aspects of RG flow and integrability in holographic duals, bridging the gap between two-dimensional $T\bar{T}$-deformed field theories and four-dimensional gravitational dynamics. The deformation parameter introduced by the root-$T\bar{T}$ flow regulates the nonlinear strength of the electromagnetic sector, preserving causality and duality at all energy scales. In this sense, the root-$T\bar{T}$ deformation provides a geometric interpretation of the flow of the Lagrangian under marginal and irrelevant couplings, unifying two-dimensional integrability with four-dimensional gravitational consistency.

\section*{Acknowledgments}
We are very grateful to Bin Chen, Dmitri Sorokin, Roberto Tateo, Jorge G. Russo, and Shahin Sheikh-Jabbari for their interest in this work and fruitful discussion.
H.B.-A. would like to express my sincere gratitude to Karapet Mkrtchyan, Neil Lambert and Alessandro Tomasiello for the highly productive discussions during the "Journey through Modern Explorations in QFT and beyond - 2 $\otimes$ 25" conference in Yerevan (August 3-13).
The work of H.B.-A. was conducted as part of the PostDoc Program on {\it Exploring TT-bar Deformations: Quantum Field Theory and Applications}, sponsored by Ningbo University.  SH would like to appreciate the financial support from Ningbo University, the Max Planck Partner group, and the Natural Science Foundation of China Grants (Nos. 12475053 and 12235016). This work is based upon research funded by Iran National Science Foundation (INSF) under project No. 4047785.

\bibliography{refs}




\end{document}